\documentclass[11pt]{article} 
\usepackage{graphicx,amssymb,mathrsfs,array,multirow}  
\begin{document}  
 
\vskip 30pt

\begin{center}  
{\Large{\bf A neutrino mass model with S3 symmetry and 
see-saw interplay}}\\
\vspace*{1cm}  
\renewcommand{\thefootnote}{\fnsymbol{footnote}}  
{ {\sf Soumita Pramanick${}$\footnote{email: soumitapramanick5@gmail.com}},
{\sf Amitava Raychaudhuri${}$\footnote{email: palitprof@gmail.com}} 
} \\  
\vspace{10pt}  
{\small  {\em Department of Physics, University of Calcutta,  
92 Acharya Prafulla Chandra Road, Kolkata 700009, India}}
\normalsize

\end{center}  

\begin{abstract} 

{\it We develop a see-saw model for neutrino masses and mixing
with an $S3\times Z3$ symmetry.  It involves an interplay
of Type-I and Type-II see-saw contributions of which the former
is subdominant. The $S3 \times Z3$ quantum numbers of the fermion
and scalar fields are chosen such  that the Type-II
see-saw generates a mass matrix which incorporates the
atmospheric mass splitting and sets $\theta_{23} = \pi/4$. The
solar splitting and $\theta_{13}$ are absent, while the
third mixing angle can achieve any value, $\theta_{12}^0$. 
Specific choices of $\theta_{12}^0$ are of interest,
e.g., $35.3^\circ$ (tribimaximal), $45.0^\circ$  (bimaximal),
$31.7^\circ$ (golden ratio), and  $0^\circ$ (no solar mixing). The
role of the Type-I see-saw is to nudge all the above into the
range indicated by the data. The model results in novel
interrelationships between these quantities due to their common
origin, making it readily falsifiable. For example, normal
(inverted) ordering is associated with  $\theta_{23}$ in the
first (second) octant. CP-violation is controlled by phases in
the right-handed neutrino Majorana mass matrix, $M_{\nu R}$. In
their absence, only normal ordering is
admissible. When $M_{\nu R}$  is complex  the Dirac CP-phase,
$\delta$, can be large, i.e., $\sim \pm \pi/2$, and inverted
ordering is also allowed.   The preliminary results from T2K and NOVA
which favour normal ordering and $\delta \sim -\pi/2$ are
indicative, in this model, of a lightest neutrino mass of 0.05 eV or
more.}

\vskip 5pt \noindent  
\texttt{Key Words:~~Neutrino mixing, $\theta_{13}$, Solar
splitting, S3, see-saw, Leptonic CP-violation}
\end{abstract}  

\renewcommand{\thesection}{\Roman{section}} 
\setcounter{footnote}{0} 
\renewcommand{\thefootnote}{\arabic{footnote}} 
\noindent

\section{Introduction}
Oscillation experiments over vastly different baselines
and a range of neutrino energies have filled up a vast portion of
the mass and mixing jigsaw of the neutrino sector. Yet, we still
remain in the dark with regard to CP-violation in the lepton
sector. Neither do we know the mass ordering -- whether it is
normal or inverted.  Further open issues are the absolute mass scale
of neutrinos and whether they are of Majorana or Dirac nature.
While we await experimental guidance for each of the
above unknowns, there have been many attempts to build 
models of lepton mass which capture much of what is known.

Here we propose a neutrino mass model based on the direct product
group $S3 \times Z3$. The elements of $S3$ correspond to the
permutations of three objects\footnote{More details of $S3$
can be found in Appendix \ref{AppS3}.}. Needless to say,
$S3$-based models of neutrino mass have been considered earlier
\cite{S3older, S3ma}. A popular point of view \cite{Idemo} has been to
note that a permutation symmetry between the three neutrino
states is consistent with\footnote{Note, however, there is no 3-dimensional irreducible 
representation of $S3$ (see Appendix \ref{AppS3}). So these models entail fine tuning.} (a) a democratic mass matrix,
$M_{dem}$, all whose elements are equal, and (b) a mass matrix
proportional to the identity matrix, $I$. A general combination
of these two forms, e.g., $c_1 I + c_2 M_{dem}$, where $c_1, c_2$
are complex numbers, provides a natural starting point. One of
the eigenstates, namely, an equal weighted combination of the
three states, is one column of the popular tribimaximal mixing
matrix. Many models have been presented \cite{Idemo} which add
perturbations to this structure to accomplish realistic neutrino
masses and mixing. Variations on this theme \cite{Idem2} explore
mass matrices with such a form in the context of Grand
Unified Theories, in models of extra dimensions, and
examine renormalisation group effects on such a pattern realised at a
high energy. Other variants of the $S3$-based models, for example
\cite{S3other}, rely on a 3-3-1 local gauge symmetry, tie it to a
$(B-L)$-extended model, or realise specific forms of mass matrices
through soft symmetry breaking, etc.    As discussed later, the
irreducible representations of $S3$ are one and two-dimensional.
The latter provides a natural mechanism to get maximal mixing in
the $\nu_\mu-\nu_\tau$ sector \cite{S3mutau}.  

The present work, also based on $S3$ symmetry, breaks new
ground in the following directions.  Firstly, it involves an
interplay of Type-I and Type-II see-saw contributions. Secondly,
it presents a general framework encompassing many popular mixing
patterns such as tribimaximal mixing. Further, 
the model does not invoke any soft symmetry breaking terms. All
the symmetries are broken spontaneously.

We briefly outline here the strategy of this work. We use
the standard notation for the leptonic mixing matrix -- the
Pontecorvo, Maki, Nakagawa, Sakata (PMNS) matrix -- $U$.
\begin{eqnarray}
U = \left(
          \begin{array}{ccc}
          c_{12}c_{13} & s_{12}c_{13} & s_{13}e^{-i\delta}  \\
- c_{23}s_{12} + s_{23}s_{13}c_{12}e^{i\delta} & c_{23}c_{12} +
s_{23}s_{13}s_{12}e^{i\delta}&  s_{23}c_{13}\\
 s_{23}s_{12} + c_{23}s_{13}c_{12}e^{i\delta}&  -s_{23}c_{12} +
c_{23}s_{13}s_{12}e^{i\delta} & c_{23}c_{13} \end{array} \right)
\;\;,
\label{PMNS}
\end{eqnarray}
where $c_{ij} = \cos \theta_{ij}$ and $s_{ij} = \sin
\theta_{ij}$.  The neutrino masses and mixings arise through a two-stage
mechanism.  In the first step, from the Type-II see-saw the
larger atmospheric mass splitting, $\Delta m^2_{atmos}$, is
generated while the solar splitting, $\Delta m^2_{solar}$,  is
absent.  Also, $\theta_{13}= 0$, $\theta_{23}= \pi/4$ and the
model parameters can be continuously varied to obtain any desired
$\theta_{12}^0$.  Of course, in reality $\theta_{13} \neq 0$
\cite{t13}, the solar splitting is non-zero, and there are
indications that $\theta_{23}$ is large but non-maximal.
Experiments have also set limits on $\theta_{12}$. The Type-I
see-saw addresses all the above issues and relates the masses and
mixings to each other.

The starting form incorporates several well-studied mixing
patterns such as tribimaximal (TBM), bimaximal (BM), and golden
ratio (GR) mixings within its fold. These alternatives all have
$\theta_{13} = 0$  and $\theta_{23} = \pi/4$. They differ only in
the value of the third mixing angle $\theta_{12}^0$ as displayed
in Table \ref{t1}. The fourth option in this Table,  no solar mixing
(NSM), exhibits the attractive feature\footnote{Such a
mixing pattern was envisioned earlier in a model based on  $A4$
symmetry \cite{OurA4} which built on previous work along similar lines
\cite{pr, br}.} that the mixing angles are
either maximal, i.e.,  $\pi/4$ ($\theta_{23}$) or vanishing
($\theta_{13}$ and $\theta_{12}^0$).

\begin{table}[tb]
\begin{center}
\begin{tabular}{|c|c|c|c|c|}
\hline
  Model &TBM &BM & GR & NSM\\ \hline
$\theta^0_{12}$ & 35.3$^\circ$ & 45.0$^\circ$  & 31.7$^\circ$ &
0.0$^\circ$ \\ \hline
\end{tabular}
\end{center}
\caption{The solar mixing angle, $\theta^0_{12}$ for this
work, for
the TBM, BM, and GR  mixing patterns. NSM stands for the case
where the solar mixing angle is initially vanishing. }
\label{t1}
\end{table}

In the following section we furnish  a description of the model
including the assignment of $S3 \times Z3$ quantum numbers to the
leptons and symmetry-breaking scalar fields. The consequences of
the model are described next where we also compare with the
experimental data. A summary and conclusions follow.  The scalar
potential of this model has a rich structure.  In two Appendices
we present the essence of $S3$ symmetry and discuss the $S3$
invariant scalar potential, deriving the conditions which must
be satisfied by the scalar coefficients to obtain the desired
minimum.

\section{The Model}
\label{Model}

In the model under discussion fermion and scalar multiplets
are assigned $S3 \times Z3$ quantum numbers in a manner such that
spontaneous symmetry breaking naturally yields mass matrices
which lead to the see-saw features espoused earlier. All terms
allowed by the symmetries of the model are included in the
Lagrangian. No soft symmetry-breaking terms are required.

To begin it will be useful to formulate the conceptual structure behind the
model. Neutrino masses arise from a combination of Type-I and
Type-II see-saw contributions of which the latter dominates. In
the neutrino mass basis, which is also the basis in which the
Lagrangian will be presented, the Type-II see-saw yields a
diagonal matrix in which two states are degenerate: 
\begin{equation}
M_{\nu L} = 
\pmatrix{m^{(0)}_1 & 0 & 0 \cr 0 & m^{(0)}_1 & 0 \cr 0 & 0 & m^{(0)}_3} \;\;.
\label{mnu1}
\end{equation}
This mass matrix results in $\Delta m^2_{atmos} = (m_3^{(0)})^2 -
(m_1^{(0)})^2$ while $\Delta m^2_{solar} = 0$.  Later, we find
the combinations $m^{\pm} = m_3^{(0)} \pm m_1^{(0)}$ useful.
$m^-$ signals the mass ordering; it is positive for normal
ordering (NO) and negative for inverted ordering (IO).

\begin{table}[t]
\begin{center}
\begin{tabular}{|c|c|c|c|c|}
\hline
Fields & Notations & $S3$ ($Z3$) & $SU(2)_L (Y)$ & $L$    \\ 
 \hline
 & & &  &   \\
& $L_e$ & $1' ~(1)$ &  &  \\ 
Left-handed leptons&$L_\mu$&$1' ~(\omega)$ & 2 (-1) & +1  \\
& $L_\tau$ & 1 ($\omega$) & & \\ 
 & & &   &  \\
\hline
 & & &  &   \\
& $e_R$ & $1' ~(1)$ &  &  \\ 
Right-handed charged leptons&$\pmatrix{\mu_R \cr \tau_R}$&$2
~(1)$ & 1 (-2) & +1 \\ & & &   &  \\
\hline
 & & &  &   \\
& $N_{1R}$ & $1' ~(1)$ &  &  \\ 
Right-handed neutrinos&$N_{2R}$&$1' ~(\omega)$ & 1 (0) & 0  \\
& $N_{3R}$ & 1 ($\omega$) & & \\ 
 & & &   &  \\
\hline
\end{tabular}
\end{center}
\caption{\em The fermion content of the model. The transformation
properties under $S3$, $Z3$, and $SU(2)_L$ are shown. The
hypercharge of the fields, $Y$, and their lepton number, $L$, are
also indicated. Here $L_\alpha^T = (\nu_\alpha \;\; l_\alpha^-)$.}
\label{tab1f}
\end{table}

At this stage the mixing resides entirely in the charged lepton
sector. We follow the convention
\begin{equation}
\Psi_{flavour} = U_{\Psi} \Psi_{mass} \;,
\end{equation}
for the fermions $\Psi$, so that the PMNS matrix, $U$, is given by  
\begin{equation}
U = U_{l}^\dagger U_\nu \;.
\label{mixmat}
\end{equation}
As noted, at this level  $\theta_{12} = \theta_{12}^0$,  where
alternate choices of $\theta_{12}^0$ result in popular mixing
patterns such as tribimaximal, bimaximal, and golden ratio with
the common feature that  $\theta_{13} = 0$ and $\theta_{23} =
\pi/4$. $\theta_{12}^0 = 0$ is another interesting alternative
\cite{OurA4} where initially the lepton mixing angles are either
vanishing $\theta_{13} = 0 = \theta_{12}$ or maximal, i.e.,
$\pi/4$ ($\theta_{23}$). Thus, till Type-I see-saw effects are
included, the leptonic mixing matrix takes the form:
\begin{equation}
U^0=
\pmatrix{\cos \theta_{12}^0 & \sin \theta_{12}^0  & 0 \cr -\frac{\sin
\theta_{12}^0}{\sqrt{2}} & \frac{\cos \theta_{12}^0}{\sqrt{2}} &
{1\over\sqrt{2}} \cr
\frac{\sin \theta_{12}^0}{\sqrt{2}} & -\frac{\cos
\theta_{12}^0}{\sqrt{2}}  & {1\over\sqrt{2}}} = U_l^\dagger U_\nu^0 \;,
\label{mix0}
\end{equation}
where $U_\nu^0 = I$ and the charged lepton mass matrix is:
\begin{equation}
M_{e\mu\tau}= U_l \pmatrix{m_e & 0 & 0 \cr 0 &
m_\mu & 0 \cr 0 & 0 & m_\tau} I =
\pmatrix{m_e \cos \theta_{12}^0 & -{m_\mu \over \sqrt{2}} \sin
\theta_{12}^0 &  {m_\tau \over \sqrt{2}} \sin \theta_{12}^0 \cr
m_e \sin \theta_{12}^0 & {m_\mu \over \sqrt{2}} \cos
\theta_{12}^0 &  -{m_\tau \over \sqrt{2}} \cos \theta_{12}^0\cr
0 & {m_\mu \over \sqrt{2}} &  {m_\tau \over \sqrt{2}}} \;\;.
\label{mcharged}
\end{equation}
The identity matrix, $I$, at the right in the first step
above indicates that no transformation needs to be applied on
the right-handed charged leptons which are $SU(2)_L$ singlets.

In this basis, the matrices responsible for the Type-I see-saw
have the forms:
\begin{equation}
M_D = m_D ~\mathbb{I} \;\; {\rm and} \;\; 
M_{\nu R} = {m_R \over 2 x y} \pmatrix{0 & x e^{-i\phi_1} & x
e^{-i\phi_1} \cr x e^{-i\phi_1} & y e^{-i\phi_2}/\sqrt{2} &  -y
e^{-i\phi_2}/\sqrt{2} \cr x e^{-i\phi_1} & -y
e^{-i\phi_2}/\sqrt{2} & y e^{-i\phi_2}/\sqrt{2}}
\;\;,
\label{mnu2}
\end{equation}
where $m_D$ and $m_R$ set the scale for the Dirac and
right-handed Majorana masses while $x$ and  $y$ are dimensionless
real quantities of $\cal O$(1). We take the Dirac mass matrix
$M_D$ proportional to the identity for ease of presentation.  We
have checked that the same results can be reproduced so long as
$M_D$ is diagonal. The right-handed neutrino Majorana mass
matrix, $M_{\nu R}$, has a $N_{2R} \leftrightarrow N_{3R}$
discrete symmetry.  This choice too can be relaxed without
jeopardising the final outcome.

We will show later how the mass matrices in eqs. (\ref{mnu1})
- (\ref{mnu2}) lead to a good fit to the neutrino data and
yield testable predictions. But before this
we must ensure that the above matrices can arise from the $S3
\times Z3$ symmetric Lagrangian. 

The behaviour of the fermions, i.e., the three lepton
generations\footnote{The scope of this model is restricted to the
lepton sector.} including three right-handed neutrinos,  is
summarised in Table \ref{tab1f}.  The gauge interactions of the
leptons are universal and diagonal in this basis. 
A feature worth noting is that the right-handed neutrinos have
lepton number $L=0$. We discuss later how this leads to a
diagonal neutrino Dirac mass matrix. The lepton mass matrices
arise from the Yukawa couplings allowed by the $S3 \times Z3$ symmetry.

\begin{table}[t]
\begin{center}
\begin{tabular}{|c|c|c|c|c|c|}
\hline
Purpose & Notations & $S3$ & $SU(2)_L$ & $L$ & $vev$ \\ 
 &  & ($Z3$) & ($Y$) & &  \\  \hline
 & & & & &  \\
& $\eta \equiv (\eta^+ ~\eta^0)$ & 1 (1) & 2 (1)
& 0 & $\langle \eta \rangle = v_\eta(0 ~1)$  \\
 & & & & &  \\
& $\Phi_a \equiv \pmatrix{\phi_1^+ & \phi_1^0\cr
\phi_2^+& \phi_2^0}$&2 (1)&2 (1) &0&$\langle\Phi_a\rangle=
\frac{v_a}{\sqrt{2}} \pmatrix{0 & w_1 \cr 0 & w_2  }$ \\
Charged fermion mass & & & & &   \\
& $\Phi_b \equiv \pmatrix{\phi_3^+ & \phi_3^0\cr
\phi_4^+& \phi_4^0}$&2 ($\omega$)&2 (1) &0&$\langle\Phi_b\rangle=
\frac{v_b}{\sqrt{2}} \pmatrix{0 & w_3  \cr 0 & w_4  }$ \\
 & & & & &  \\
& $\alpha \equiv (\alpha^+ ~\alpha^0)$ & 1 ($\omega$) & 2 (1)
& 0 & $\langle \alpha \rangle = v_\alpha(0 ~1)$  \\
 & & & & &  \\
 \hline
 & & & & &  \\
 Neutrino Dirac mass& $\beta \equiv (\beta^0 ~\beta^- )$
&1 (1) &2 (-1)& 1& $\langle\beta\rangle=v_\beta\pmatrix{1 ~0}$
\\
 & & & & &  \\
\hline
 & & & &  & \\
 &$\Delta_L \equiv
(\Delta_L^{++}, ~\Delta_L^+, ~\Delta_L^0)$  &1 (1) &3 (2) & -2 &
$\langle\Delta_L\rangle= v_\Delta \pmatrix{0 ~0 ~1}$ \\
Type-II see-saw mass & & & & &  \\
 &$\rho_L \equiv
(\rho_L^{++}, ~\rho_L^+, ~\rho_L^0)$  &1 ($\omega$) &3 (2) & -2 &
$\langle\rho_L\rangle=v_\rho\pmatrix{0 ~0 ~1}$ \\
 & & & & &  \\
\hline
 & & & &  & \\
 &$\chi \equiv \chi^0$  & 1 ($\omega$) & 1 (0) & 0 &
$\langle\chi\rangle= u_\chi$ \\
 Right-handed neutrino mass & & & & &  \\
 &$\gamma \equiv \gamma^0$  & $1' ~(\omega)$ & 1 (0) & 0 &
$\langle\gamma\rangle= u_\gamma$ \\
 & & & & &  \\
 \hline
\end{tabular}
\end{center}
\caption{\em The scalar content of the model. The transformation
properties under $S3$, $Z3$, and $SU(2)_L$ are shown. The hypercharge
of the fields, $Y$, their lepton number, $L$, and the vacuum
expectation values are also indicated.  $w_i ~(i=1 \ldots 4)$
are dimensionless. } 
\label{tab1s}
\end{table}

The $S3 \times Z3$ structure of the lepton sector is matched by a
rich scalar sector which we have presented in Table
\ref{tab1s}. The requirement of charged lepton masses and Type-I
and Type-II see-saw neutrino masses dictates the inclusion of
$SU(2)_L$ singlet, doublet, and triplet scalar fields. The
$S3 \times Z3$ properties of the scalars are chosen bearing in
mind the $S3$ and $Z3$ combination  rules. In particular, for the
former the representations are $1$, $1'$, and $2$ which satisfy
the multiplication rules (see Appendix \ref{AppS3}):
\begin{equation}
1 \times 1^\prime = 1^\prime \;, \;\; 1^\prime \times 1^\prime = 1
\;\;, {\rm ~and} \;\; 2 \times 2 = 2 + 1 + 1^\prime \;.
\label{S3prod}
\end{equation}
The scalar multiplets are chosen such that the mass matrices
appear with specific structures as discussed below\footnote{In
general the multiple scalar fields in models based on discrete
symmetries also result in flavour changing neutral currents
induced by the neutral scalars.  Discussions of this aspect in
the context of $S3$ can be found, for example, in
\cite{S3fcnc}.}. It can be seen from Table \ref{tab1s} that all
neutral scalars pick up a {\em vev}. The {\em vev} of the
$SU(2)_L$ singlets, namely, $u_\chi$ and $u_\gamma$, can be much
higher than the electroweak scale, $v$, and determine the masses
of the right-handed neutrinos. The other {\em vev} break
$SU(2)_L$. We take $v_\Delta \sim v_\rho \ll v_\eta \sim v_a \sim v_b
\sim v_\alpha \sim v_\beta \sim v$.

Charged lepton and neutrino masses are obtained from the
Yukawa terms in a Lagrangian constructed out of the fields in
Tables \ref{tab1f} and \ref{tab1s}. Including all terms which
respect the $SU(2)_L \times 
U(1)_Y$ gauge symmetry and the $S3 \times
Z3$ flavour symmetry so long as lepton number, $L$, is also
conserved one is led to the Lagrangian mass terms
\begin{eqnarray}
\mathscr{L}_{mass}&=& f_1 ~\bar{e}_{L} (\mu_{R}\phi_2^{0} -
\tau_{R}\phi_1^{0}) + f_2 ~\bar{\mu}_{L} (\mu_{R}\phi_4^{0} -
\tau_{R}\phi_3^{0}) + f_3 ~\bar{\tau}_{L} (\mu_{R}\phi_4^{0} +
\tau_{R}\phi_3^{0}) \nonumber\\
&+&  f_4 ~\bar{\mu}_{L} e_{R}\alpha^{0} +  f_5
~\bar{e}_{L} e_{R}\eta^{0}  
 ~~{\rm(charged ~lepton ~mass)}  \nonumber\\
&+& (h_1 \bar{\nu}_{eL} N_{1R} + h_2 \bar{\nu}_{\mu L} N_{2R} + h_3
\bar{\nu}_{\tau L} N_{3R}) \beta^0 
 ~~{\rm(neutrino ~Dirac ~mass)} \nonumber\\
&+&\frac{1}{2} g_1 ~\nu_{eL}^TC^{-1}\nu_{eL} \Delta_L^0 + 
\frac{1}{2} \left(g_2 ~\nu_{\mu L}^TC^{-1}\nu_{\mu L} + 
g_3 ~\nu_{\tau L}^TC^{-1}\nu_{\tau L} \right) \rho_L 
 ~~{\rm(neutrino ~Type\!-\!II ~see\!-\!saw ~mass)} \nonumber \\
&+& \frac{1}{2}\left( \left[k_1  N_{2R}^TC^{-1}N_{2R} + k_2
N_{3R}^TC^{-1}N_{3R}\right] \chi + k_3  N_{2R}^TC^{-1}N_{3R}
\gamma \right) \nonumber \\
&+&  \frac{1}{2}\left(k_4  N_{1R}^TC^{-1}N_{2R} \tilde{\chi} + k_5
N_{1R}^TC^{-1}N_{3R} \tilde{\gamma} \right) ~~{\rm(rh ~neutrino
~mass)} + h.c. \;\;.
\label{e1}
\end{eqnarray}
Here, $\tilde{\chi}$ and $\tilde{\gamma}$ are charge
conjugated fields which transform under $Z_3$ as $\omega^2$. For
each term in the Lagrangian  the fermion masses which arise
therefrom have been indicated.  Both Type-I and Type-II see-saw
contributions for neutrino masses are present. 

The above Lagrangian gives rise to the mass matrices in
Eqs.  (\ref{mnu1}) - (\ref{mnu2}) through the Yukawa couplings in
Eq. (\ref{e1}) and the {\em vev}s in Table \ref{tab1s}. Before
turning to these let us note how the quantum number assignments
of the fermion and scalar fields force certain entries in the
mass matrices to be vanishing.  For example, the mass term
$\bar{\tau}_L e_R$ is zero in Eq.  (\ref{mcharged}) because there
is no $SU(2)_L$ doublet field which transforms as a $1'$ under
$S3$. Similarly the diagonal nature of the left-handed neutrino
Majorana mass matrix in Eq. (\ref{mnu1}) is ensured by the
absence of an $SU(2)_L$ triplet field which transforms either as
(i) a $1'$ under $S3$ or (ii) as $\omega^2$ under $Z_3$. The
neutrino Dirac mass matrix in Eq. (\ref{mnu2}) arises from the
Yukawa couplings\footnote{As the $N_{iR}$ carry $L = 0$,
conservation of lepton number forbids any contribution to the
Dirac mass from the $SU(2)_L$  scalar doublets which generate the
charged lepton masses.} of the $SU(2)_L$ doublet scalar $\beta$.
Since it transforms as 1 under both $S3$ and $Z3$ it can be seen
from the left-handed and right-handed neutrino quantum numbers in
Table \ref{tab1f} that only diagonal terms are allowed. Finally,
the $N_{1R}^T N_{1R}$ term is absent in the right-handed neutrino
Majorana mass matrix in Eq.  (\ref{mnu2}) since there is no $Z3$
singlet among the $SU(2)_L$ singlet scalars.

Before proceeding further it may be useful to comment on
the sizes of the various vacuum expectation values in Table
\ref{tab1s}. The $SU(2)_L$ doublets acquire {\em vev}s $v_{\eta,
a, b, \alpha, \beta}$ which are ${\cal O}(M_W)$ while the triplet
{\em vev}s $v_{\Delta, \rho}$ are several orders of magnitude
smaller. This is in consonance with the smallness of the neutrino
masses as also the $\rho$ parameter of electroweak symmetry
breaking. Needless to say, the $SU(2)_L$ singlet fields $\chi$
and $\gamma$ can acquire {\em vev}s well above the electroweak
scale.

The non-vanishing entries in the mass matrices in Eqs.
(\ref{mnu1}) - (\ref{mnu2}) which arise from the Yukawa
couplings entail the following relationships:

1. Charged lepton masses -- On matching the Lagrangian in Eq.
(\ref{e1}), the scalar doublet {\em vev}s in Table \ref{tab1s} and
the charged lepton mass matrix in Eq. (\ref{mcharged}) one gets:  
\begin{equation}
f_1 \langle\phi_1^0\rangle = -\frac{m_\tau}{\sqrt{2}} \sin
\theta_{12}^0 \;\;,\;\;f_1 \langle\phi_2^0\rangle =
-\frac{m_\mu}{\sqrt{2}} \sin
\theta_{12}^0 \;\;,
\label{mc1}
\end{equation}
\begin{equation}
f_2 \langle\phi_3^0\rangle = \frac{m_\tau}{\sqrt{2}} \cos
\theta_{12}^0 \;\;,\;\;f_2 \langle\phi_4^0\rangle = \frac{m_\mu}{\sqrt{2}} \cos
\theta_{12}^0 \;\;,\;\;
f_3 \langle\phi_3^0\rangle = \frac{m_\tau}{\sqrt{2}} 
\;\;,\;\;f_3 \langle\phi_4^0\rangle = \frac{m_\mu}{\sqrt{2}} \;\;,
\label{mc2}
\end{equation}
and
\begin{equation}
f_4 \langle\alpha^0\rangle = m_e \sin\theta_{12}^0 \;\;,\;\;
f_5 \langle\eta^0\rangle = m_e \cos\theta_{12}^0 \;\;.
\label{mc3}
\end{equation}
Notice that Eqs. (\ref{mc1}) and (\ref{mc2}) imply 
\begin{equation}
\frac{w_2}{w_1} = \frac{w_4}{w_3} = \frac{m_\mu}{m_\tau}  \;\;.
\label{vevrat}
\end{equation}

2. Left-handed neutrino Majorana mass -- Similarly, the mass
matrix in Eq. (\ref{mnu1}) is obtained when
\begin{equation}
g_1 \langle\Delta_L^0\rangle = m_1^0 = g_2 \langle\rho_L^0\rangle  \;\;,\;\;
g_3 \langle\rho_L^0\rangle = m_3^0 
\;\;.
\label{mn1}
\end{equation}
The first equation above requires a matching between two
sets of Yukawa couplings and {\em vev}s. This is to
ensure degeneracy of two neutrino states, implying the vanishing of
the solar mass splitting at this stage. Notice that the
relatively large size of the atmospheric mass splitting requires
$g_2$ and $g_3$ to be of different order.

3. Neutrino Dirac mass -- The Dirac mass matrix in Eq.
(\ref{mnu2}) is due to the relations: 
\begin{equation}
h_1 = h_2 = h_3 = h   \;\; {\rm and} \;\;
h \langle\beta^0\rangle = m_D 
\;\;.
\label{mn2}
\end{equation}
The equality of the three Yukawa couplings, $h_i$, above is
only a simplified choice. We have checked that deviations from
this relation, i.e., a diagonal Dirac mass matrix but not
proportional to the identity, can also readily lead to the
results which we discuss in this paper.

4. Right-handed neutrino Majorana mass -- Finally, the
right-handed neutrino Majorana mass matrix follows from:
\begin{equation}
k_1 \langle \chi^0
\rangle = \frac{m_R e^{-i\phi_2}}{2\sqrt{2}x} = k_2 \langle \chi^0 \rangle 
\;\;,\;\; k_3 \langle \gamma^0 \rangle = -\frac{m_Re^{-i\phi_2}}{2\sqrt{2}x}
\;\;,\;\; k_4 \langle\tilde{\chi}^0\rangle = \frac{m_R e^{-i\phi_1}}{2y} = k_5
\langle\tilde{\gamma}^0\rangle  
\;\;.
\label{mn3}
\end{equation}
We show in Appendix \ref{AppPot} how from a minimisation of the scalar
potential the required scalar {\em vev}s may be obtained.

\subsection{Type-I see-saw contribution}

In the previous section we have shown that the $S3$ model
results in a diagonal left-handed neutrino mass matrix
given in Eq. (\ref{mnu1}) through a Type-II see-saw. The charged
lepton mass matrix as given in Eq.
(\ref{mcharged}) is not diagonal and induces a mixing in the
lepton sector. This mixing, Eq.  (\ref{mix0}), receives further
corrections from a smaller Type-I see-saw contribution to the neutrino
mass matrix as we discuss.

The Type-I see-saw arising from the Dirac and right-handed
neutrino mass matrices in Eq. (\ref{mnu2}) is 
\begin{equation}
 M' =  \left[M_D^T(M_{\nu R})^{-1}M_D \right]  
= {m_D^2 \over m_R} 
\pmatrix{0 & y~e^{i\phi_1} & y~e^{i\phi_1} \cr 
y~e^{i\phi_1} & {x~e^{i\phi_2} \over  \sqrt 2} 
& {-x~e^{i\phi_2} \over \sqrt 2} \cr 
y~e^{i\phi_1} & {-x~e^{i\phi_2} \over \sqrt 2} & 
{x~e^{i\phi_2} \over \sqrt 2}}\;\;.
\label{pertmat}
\end{equation}

\section{Results}

We have given above the contributions to the neutrino
mass matrix from the Type-I and Type-II see-saw. Of these, the
former is taken to be significantly smaller than the latter. As
we have noted, in the absence of the Type-I see-saw the leptonic
mixing matrix in this model is determined entirely by the charged
lepton mass matrix. It has $\theta_{13} = 0$, $\theta_{23} =
\pi/4$, and $\theta_{12}$ arbitrary. We will be considering four
mixing patterns which fall within this scheme  and in
each of which the value of $\theta_{12}^0$ is specified, namely, the
TBM, BM, GR, and NSM cases. In addition, in this model the
Type-II see-saw sets the solar mass splitting to be zero. The
Type-I see-saw, whose effect we incorporate perturbatively, brings
all the above leptonic parameters into agreement with their
values preferred by the data.  Before we proceed further with
this discussion it will be useful to summarise the global best-fit
values of these mass-splittings and angles.

\subsection{Data}
\label{sec:data}

From global fits the currently favoured 3$\sigma$ ranges of the neutrino
mixing parameters are
\cite{Gonzalez, Valle} 
\begin{eqnarray}
\Delta m_{21}^2 &=& (7.02 - 8.08) \times 10^{-5} \, {\rm eV}^2, \;\;
\theta_{12} = (31.52 - 36.18)^\circ, \nonumber \\
|\Delta m_{31}^2| &=& (2.351 -  2.618) \times 10^{-3}
\, {\rm eV}^2, \;\;
\theta_{23} = (38.6 - 53.1)^\circ \,, \nonumber \\
\theta_{13} &=& (7.86 - 9.11)^\circ, \;\; \delta =
(0 - 360)^\circ \;\;.
\label{results}
\end{eqnarray}
These data are from NuFIT2.1 of 2016 \cite{Gonzalez}. Here,
$\Delta m_{ij}^2 = m_i^2 - m_j^2$, so that $\Delta m_{31}^2 > 0
~(<0)$ for normal (inverted) ordering. The data indicate two
best-fit points for $\theta_{23}$ in the first and second
octants. Later, we also remark about the compatibility of
this model with the recent  T2K and NOVA hints
\cite{T2K, Nova} of $\delta$ being near -$\pi/2$.

\subsection{Real $M_{\nu R}$ ($\phi_1 = 0
~{\rm or} ~\pi, \phi_2 = 0 ~{\rm or} ~\pi$)}\label{sec3}

A limiting case, with less complications, corresponds to no
CP-violation.  This happens when $M_{\nu R}$ is real, i.e., the phases
$\phi_{1,2}$ in Eq.  (\ref{pertmat}) are 0 or $\pi$.  These
cases can be compactly considered by keeping $x$ and $y$ real but
allowing them to be of either sign, i.e., four alternatives. We
show below how the experimental data picks out one or the other out of these.

Without the phases $\phi_{1,2}$, i.e., for real $M_{\nu R}$, one gets
\begin{equation}
 M' = 
{m_D^2 \over  m_R} 
\pmatrix{0 & y & y \cr y & {x \over  \sqrt 2} & -{x \over \sqrt 2} \cr 
y & -{x \over \sqrt 2} & {x \over \sqrt 2}}\;\;.
\label{pert1}
\end{equation}

The equality of two neutrino masses from the Type-II see-saw
requires the use of degenerate perturbation theory to obtain
corrections  to the solar mixing parameters.  The 
$2\times2$ submatrix of $M'$ relevant for this is:
\begin{equation}
M'_{2\times2} = {m_D^2 \over  m_R} 
\pmatrix{0 & y \cr y & {x/\sqrt 2}} \;.
\label{solr}
\end{equation}
This results in:
\begin{equation}
\theta_{12} = \theta_{12}^0 + \zeta \;\;,\;\; \tan 2\zeta= 2 \sqrt 2
\left(\frac{y}{x}\right)  \; .
\label{solangr}
\end{equation}
\begin{table}[tb]
\begin{center}
\begin{tabular}{|c|c|c|c|c|}
\hline
Model ($\theta^0_{12}$) &TBM (35.3$^\circ$) &BM (45.0$^\circ$) &
GR (31.7$^\circ$)  & NSM (0.0$^\circ$) \\ \hline
$\zeta$  & -4.0$^\circ \leftrightarrow$ 0.6$^\circ$ & -13.7$^\circ
\leftrightarrow$  -9.1$^\circ$  & -0.4$^\circ \leftrightarrow$
4.2$^\circ$ & 31.3$^\circ \leftrightarrow$  35.9$^\circ$ \\ \hline 
$\epsilon$  & -4.0$^\circ \leftrightarrow$  0.6$^\circ$ &
-14.5$^\circ \leftrightarrow$  -9.3$^\circ$  & -0.4$^\circ
\leftrightarrow$ 4.2$^\circ$ & 44.0$^\circ \leftrightarrow$
56.7$^\circ$ \\ \hline 
$\epsilon -  \theta_{12}^0$ & -39.2$^\circ
\leftrightarrow$  -34.6$^\circ$ & -59.5$^\circ
\leftrightarrow$  -54.4$^\circ$  & -39.2$^\circ \leftrightarrow$
-30.0$^\circ$ & 44.0$^\circ \leftrightarrow$  56.7$^\circ$ \\
\hline
\end{tabular}
\end{center}
\caption{The ranges of $\zeta$
(Eq. (\ref{solangr})), $\epsilon$ (Eq. (\ref{etar})), and $(\epsilon -
\theta_{12}^0)$ allowed by the data for the different popular
mixing patterns. }
\label{tlim}
\end{table}
A related quantity, $\epsilon$, which is found useful later is
given by
\begin{equation}
\sin\epsilon = \frac{y}{\sqrt{y^2 + x^2/2}}\;\;{\rm and} ~\cos\epsilon
= \frac{x/\sqrt{2}}{\sqrt{y^2 + x^2/2}}\;\;,\;\;{\rm i.e.,} ~\tan \epsilon =
\frac{1}{2} \tan 2\zeta \;.
\label{etar}
\end{equation} 
Once a mixing pattern is chosen, i.e.,
$\theta_{12}^0$ fixed, the experimental limits on $\theta_{12}$
as given in Eq.  (\ref{results}) set bounds on the range of
$\zeta$ and also from Eq. (\ref{etar}) on $\epsilon$. These are
displayed for the four mixing patterns in Table \ref{tlim}. If
$\zeta$ is positive (negative) then the ratio $(y/x)$ will also
be positive (negative). In addition, from Eq. (\ref{etar}) the
sign of $y$ is fixed by the value of $\epsilon$. Taking these
points into account one can conclude that $x$ is always positive,
i.e., $\phi_2$  has to be 0, while $y$ must be
positive, $\phi_1 = 0$ (negative, $\phi_1 = \pi$) for NSM (BM).
For the other mixing patterns, i.e., TBM and GR, both signs of $y$
are possible.

The solar mass splitting arising from the Type-I see-saw is
also obtained from Eq. (\ref{solr}).
\begin{equation}
\Delta m^2_{solar}=  {\sqrt{2} m_D^2 \over m_R} ~m^{(0)}_1
\sqrt{x^2 + 8y^2} =  {\sqrt{2} m_D^2 \over m_R} ~m^{(0)}_1
\frac{x}{\cos 2\zeta} \;\;. 
\label{solspltr}
\end{equation} 
Furthermore, incorporating the leading order corrections to
neutrino mixing from Eq. (\ref{pert1}) one gets from Eq. (\ref{mixmat}):
\begin{equation}
U = U^0 U_\nu \; {\rm with} ~U_\nu=
\pmatrix{\cos \zeta & -\sin \zeta  & \kappa_r \sin\epsilon \cr 
\sin\zeta & \cos \zeta & -\kappa_r \cos\epsilon \cr
\kappa_r \sin (\zeta - \epsilon) & \kappa_r \cos (\zeta - \epsilon) & 1} \;,
\label{mixr}
\end{equation}
with
\begin{equation}
\kappa_r \equiv {m_D^2 \over {m_R m^-}} \sqrt{y^2 + x^2/2} = 
{m_D^2 \over {m_R m^-}} \frac{x}{\sqrt{2} \cos\epsilon}\;\;.
\label{kappa}
\end{equation} 
The third column of the leptonic mixing matrix becomes:
\begin{equation}
|\psi_3\rangle =
\pmatrix{\kappa_r \sin (\epsilon - \theta_{12}^0) \cr 
{1\over \sqrt 2}{[1 - \kappa_r \cos (\epsilon - \theta_{12}^0)]} 
\cr {1\over \sqrt 2}{[1 + \kappa_r \cos (\epsilon - \theta_{12}^0)]} } \ \ \ .
\label{psi3_1}
\end{equation} 
Since, as noted, $x$ is always positive, $\kappa_r$ is positive
(negative)  for normal (inverted) ordering. 

The right-hand-side of Eq. (\ref{psi3_1}) has to be matched with
the third column of eq. (\ref{PMNS}). This yields:
\begin{equation}
\sin \theta_{13}\cos\delta = \kappa_r \sin (\epsilon -
\theta_{12}^0) \;\;, 
\label{s13r}
\end{equation}
and 
\begin{equation}
\tan (\pi/4 - \theta_{23}) \equiv \tan \omega = \kappa_r \cos
(\epsilon - \theta_{12}^0) \;.
\label{omegar}
\end{equation}
For ready reference, the ranges of $(\epsilon - \theta_{12}^0)$
allowed for the different mixing patterns are presented in Table
\ref{tlim}. For normal ordering\footnote{We show in the following
that inverted ordering is not consistent with real $M_{\nu R}$.}
the CP-phase $\delta$ is 0 ($\pi$) when $\sin (\epsilon - \theta_{12}^0)$ is
positive (negative). From Table \ref{tlim} one can then observe
that $\delta = 0$ for the NSM mixing pattern and is $\pi$ for the
three other cases. Needless to say, both correspond to CP-conservation.

\begin{figure}[tbh]
\begin{center}
\includegraphics[scale=0.75,angle=0]{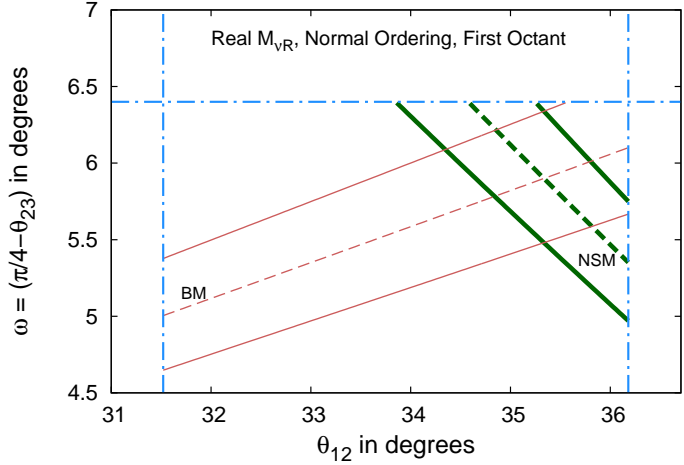}
\end{center}
\caption{\em  $\omega = (\pi/4 - \theta_{23})$
as a function of $\theta_{12}$ for normal ordering. The solid
lines indicate the range for the $3 \sigma$ allowed variation of
$\sin \theta_{13}$ while the dashed line corresponds to the
best-fit value. Thick green (thin pink) lines are for the NSM 
  (BM) case. The horizontal and vertical blue dot-dashed lines delimit the
3$\sigma$ allowed range from data. Note that $\omega$ is always
positive, i.e., the first octant of $\theta_{23}$ is preferred.
For the TBM and GR mixing patterns $\omega$, still positive,  lies beyond
the $3\sigma$ range.  Best-fit values of the solar and
atmospheric splittings are used.  For $M_{\nu R}$ real there is no
allowed solution for inverted ordering.}
\label{Real} 
\end{figure} 

Combining Eqs. (\ref{solspltr}), (\ref{kappa}),  and
(\ref{s13r}) one can write:
\begin{equation}
\Delta m^2_{solar} =  2~m^- m^{(0)}_1 
~\frac{\sin \theta_{13} \cos\delta \cos \epsilon}{\cos 2\zeta ~\sin
(\epsilon - \theta_{12}^0)}  \;\;. 
\label{solsplr2}
\end{equation}
Eq. (\ref{solsplr2}) leads to the conclusion that inverted
ordering is not allowed for this case of real $M_{\nu R}$. 
To establish this property one can define:
\begin{equation} 
z \equiv m^- m^{(0)}_1/\Delta m^2_{atmos} \;\; {\rm and} \;\;
\tan \xi \equiv m_0/\sqrt{|\Delta m^2_{atmos}|} \;\;,
\label{zxi}
\end{equation}
where $z$ is positive for both mass orderings. From Eq.
(\ref{solsplr2})  one has
\begin{equation}
z = \left(\frac{\Delta m^2_{solar}}{|\Delta m^2_{atmos}|}\right)
\left(\frac{\cos 2\zeta ~\sin (\epsilon - \theta_{12}^0)}{2 \sin
\theta_{13} |\cos\delta| \cos \epsilon}\right) \;\;.
\label{z}
\end{equation} 
It is easy to verify from Eq. (\ref{zxi}) that 
\begin{eqnarray}
z &=& \sin \xi/(1+ \sin \xi) \;\; {\rm ~i.e.,} \;\; 0 \leq z \leq
\frac{1}{2} \;\; {\rm (for ~normal ~ordering)},\nonumber \\
z &=& 1/(1+ \sin \xi) \;\; {\rm ~i.e.,} \;\; \frac{1}{2} \leq z \leq 1 \;\;
{\rm (for  ~inverted  ~ordering)} \;\;. 
\label{m_0}
\end{eqnarray} 
There is a one-to-one correspondence of $z$ with the lightest
neutrino mass $m_0$. The quasi-degeneracy limit, i.e., $m_0
\rightarrow $ large, is
approached as $z \rightarrow \frac{1}{2}$ for both mass orderings.

In  Eq. (\ref{z})  $|\cos\delta| = 1$ for real $M_{\nu R}$. Using
the global fit mass splittings and mixing angles given in Sec.
\ref{sec:data} and Table \ref{tlim} one finds $z\sim 10^{-2}$ or
smaller for all four mixing patterns. This excludes the inverted mass
ordering option for real $M_{\nu R}$.

From Eqs. (\ref{s13r}) and (\ref{omegar}) one has
\begin{equation}
\tan \omega =  
\frac{\sin \theta_{13}\cos\delta}{\tan (\epsilon -
\theta_{12}^0)} \;\;.
\label{omegar2}
\end{equation}
The noteworthy point is that for normal ordering Eq.
(\ref{omegar}) implies that $\omega$ is always positive
irrespective of the mixing pattern. So, in this model
$\theta_{23}$ is restricted to the first octant only for real
$M_{\nu R}$.

Eqs. (\ref{solangr}) and (\ref{etar}) can be used to express
$\epsilon$ in terms of $\theta_{12}$ and thereby put $\omega$ in
Eq. (\ref{omegar2}) as a function of $\theta_{12}$ and
$\theta_{13}$ only. In Fig. \ref{Real},
$\omega$ is shown as a function of $\theta_{12}$ for the NSM
(thick green lines) and BM (thin pink lines) mixing patterns. The
ranges of $\theta_{12}$ and $\omega$ have been kept within their
3$\sigma$ allowed limits from global fits as given in Sec.
\ref{sec:data}.  The TBM and GR cases are excluded because for
the allowed values of  $\theta_{12}$ they predict $\theta_{23}$
beyond the 3$\sigma$ range.  The solid lines in the figure
correspond to the 3$\sigma$ limiting values of $\theta_{13}$  and
the dashed line is for its best-fit value. The blue dot-dashed
horizontal and vertical lines display the 3$\sigma$  experimental
bounds on $\theta_{23}$ and $\theta_{12}$. 

Using Eq. (\ref{z}) any allowed point in the $\omega - \theta_{12}$ plane
and the associated $\theta_{13}$ can be translated to a value of
$z$ or equivalently $m_0$, provided the solar and
atmospheric mass splittings  are given. We find that for both the allowed
mixing patterns the range of variation of $m_0$ is very small.
For the NSM (BM) case this range is $2.13$ meV
$\leq m_0 \leq 3.10$ meV ($3.20$ meV $\leq m_0 \leq 4.42$ meV)
when both neutrino mass splittings and all mixing angles are
varied over their full 3$\sigma$ ranges.

To summarise the real $M_{\nu R}$ case:
\begin{enumerate}
\item Only the normal mass ordering is allowed.
\item $\theta_{23}$ can lie only in the first octant.
\item The TBM and GR alternatives are inconsistent with the allowed ranges
of the neutrino mixing angles even after including the
Type-I see-saw corrections.
\item For the NSM and BM  mixing
patterns real $M_{\nu R}$  can give consistent
solutions for the neutrino masses and mixings. The ranges of
allowed lightest neutrino masses are very tiny. 
\end{enumerate}


\subsection{Complex $M_{\nu R}$}

Keeping $M_{\nu R}$ real eliminates  CP-violation.  Further,
inverted ordering is disallowed. Also, the TBM and GR mixing
patterns cannot be accommodated. These restrictions can be
ameliorated by taking $M_{\nu R}$ in its general complex form
giving rise to the Type-I see-saw contribution $M'$ as given in
Eq. (\ref{pertmat}).  Recall that this introduces the phases
$\phi_{1,2}$ and $x$ and $y$ take only positive values.  

With its complex entries, $M'$ is now not hermitian any
more. To address this we consider the combination $(M^0 +
M')^\dagger(M^0 + M')$, and treat $M^{0\dagger} M^0$ as the
leading term with $(M^{0\dagger} M' + M'^\dagger M^0)$ acting as
a perturbation at the lowest order, both hermitian by
construction. The unperturbed eigenvalues are thus
$(m^{(0)}_i)^2$. The perturbation matrix is
\begin{equation}
(M^{0\dagger} M' + M'^\dagger M^0) = 
{m_D^2 \over m_R}
\pmatrix{ 0 & 2 y m^{(0)}_1  \cos\phi_1 & 
y f(\phi_1) \cr
2 y m^{(0)}_1 \cos\phi_1 & \sqrt{2} x m^{(0)}_1  \cos\phi_2 & 
-{x \over\sqrt{2}}  f(\phi_2)\cr
y f^*(\phi_1)& -{x \over\sqrt{2}} f^*(\phi_2)& 
\sqrt{2} x m^{(0)}_3 \cos\phi_2} \;.
\label{pertcmplx}
\end{equation}
In the above
\begin{equation}
f(\varphi) = m^{+} \cos\varphi - i m^{-} \sin\varphi \;\;.
\label{ffn}
\end{equation}
The remaining calculation proceeds in much the same manner as for
real $M_{\nu R}$ while keeping the distinctive features of Eq.
(\ref{pertcmplx}) in mind.

In place of Eqs.
(\ref{solangr}) and (\ref{etar})  for the real $M_{\nu R}$ case, we get
from (\ref{pertcmplx})
\begin{equation}
\theta_{12} = \theta_{12}^0 + \zeta \;\;,\;\; \tan 2\zeta =
2\sqrt2 ~{y \over x} ~{\cos\phi_1\over\cos\phi_2} \;\;,
\label{solangcmplx}
\end{equation}
and
\begin{equation}
\sin \epsilon = \frac{y \cos\phi_1}{\sqrt{y^2 \cos^2\phi_1 + x^2
\cos^2\phi_2/2}}  \;\;,\;\;
\cos \epsilon = \frac{x \cos\phi_2/\sqrt{2}}{\sqrt{y^2 \cos^2\phi_1 + x^2
\cos^2\phi_2/2}}  \;\;,\;\;
\tan \epsilon = \frac{1}{2} \tan 2\zeta \;.
\label{etac}
\end{equation}
\begin{table}[tbh]
\begin{center}
\begin{tabular}{|c|c|c|c|c|}
\hline
Mixing & \multicolumn{2}{|c|}{Normal Ordering}  & 
\multicolumn{2}{c|}{Inverted Ordering} \\ \cline{2-5} 
Pattern & $\delta$ & $\theta_{23}$& $\delta$
& $\theta_{23}$  \\ 
& quadrant & octant & quadrant & octant \\
\hline
NSM & First/Fourth &
First & Second/Third & Second    \\ \hline
BM, TBM, GR &   Second/Third  &
First  & First/Fourth & Second  \\  \hline
\end{tabular}
\end{center}
\caption{\em Quadrants of the leptonic CP-phase $\delta$
and the octant of $\theta_{23}$ for both mass orderings for
different mixing patterns.  }
\label{tabCP}
\end{table}

The allowed ranges of $\zeta$ and $\epsilon$ depend on the mixing
pattern and are given in Table \ref{tlim}.  It is seen that for
all patterns $\cos \epsilon$ is positive. Therefore, from Eq.
(\ref{etac}) we can immediately conclude that $\phi_2$ must be
always in the first or fourth quadrants.
The possible quadrants of $\phi_1$ are also determined
from the range of $\epsilon$ for the different mixing patterns.  From
the first relation in  Eq. (\ref{etac}) we find that $\phi_1$ has
to be in the first or fourth (second or third) quadrants if
$\epsilon$ is positive (negative).   Using the results in Table
\ref{tlim} we conclude that the first (second) option is valid
for the NSM (BM) patterns. For TBM and GR cases $\epsilon$ spans a
range over positive and negative values and so both options are
included.

The solar mass splitting is induced entirely through the Type-I
see-saw contribution. From Eq. (\ref{pertcmplx}) one finds: 
\begin{equation}
\Delta m^2_{solar}
= \sqrt{2} m^{(0)}_1 ~\frac{m_D^2 }{m_R} \sqrt{x^2 \cos^2\phi_2
+ 8 y^2 \cos^2 \phi_1} 
= \sqrt{2} m^{(0)}_1 ~\frac{m_D^2 }{m_R} ~\frac{x \cos
\phi_2}{\cos 2\zeta} = \sqrt{2} m^{(0)}_1 ~\frac{m_D^2 }{m_R}
 \frac{2\sqrt{2}y \cos\phi_1}{\sin 2\zeta}
\;.
\label{solsplc1}
\end{equation}
Eq. (\ref{psi3_1}) is now replaced by: 
\begin{equation}
|\psi_3\rangle =
\pmatrix{\kappa_c [\frac{\sin\epsilon}{\cos\phi_1} f(\phi_1) \cos \theta_{12}^0 - 
\frac{\cos\epsilon}{\cos\phi_2} f(\phi_2) \sin \theta_{12}^0] /m^+ \cr 
{1\over \sqrt 2}\{1- \kappa_c [\frac{\sin\epsilon}{\cos\phi_1}
f(\phi_1) \sin \theta_{12}^0 +
\frac{\cos\epsilon}{\cos\phi_2}  f(\phi_2) \cos \theta_{12}^0]/m^+\} \cr 
{1\over \sqrt 2}\{1+ \kappa_c [\frac{\sin\epsilon}{\cos\phi_1}
f(\phi_1) \sin \theta_{12}^0 +
\frac{\cos\epsilon}{\cos\phi_2}  f(\phi_2) \cos \theta_{12}^0]/m^+\}
} \;,
\label{psi3ca}
\end{equation}
where
\begin{equation}
\kappa_c = \frac{m_D^2}{m_R m^-} ~{\sqrt{y^2 \cos^2\phi_1 + x^2
\cos^2\phi_2/2}} \;,
\end{equation}
Eq. (\ref{etac}) has been used,
and the complex function $f(\phi_{1,2})$ is defined in Eq. (\ref{ffn}).

$\kappa_c$ is positive
(negative) for normal  (inverted) ordering.
Comparing the right-hand-side of  Eq.  (\ref{psi3ca}) with the
third column of Eq. (\ref{PMNS}) we find  
\begin{equation}
\sin \theta_{13}\cos\delta = \kappa_c  ~\sin(\epsilon-
\theta_{12}^0) \;, 
\label{cdelcomp}  
\end{equation}
\begin{equation}
\sin \theta_{13}\sin\delta = \kappa_c ~\frac{m^-}{m^+
\cos\phi_1 \cos\phi_2}  
\left[\sin \epsilon \sin\phi_1 \cos \phi_2  \cos \theta_{12}^0
- \cos \epsilon \cos\phi_1 \sin \phi_2  \sin \theta_{12}^0 \right] \;.
\label{sdelcomp}
\end{equation}

\begin{figure}[tbh]%
\begin{center}
{\includegraphics[scale=0.65,angle=0]{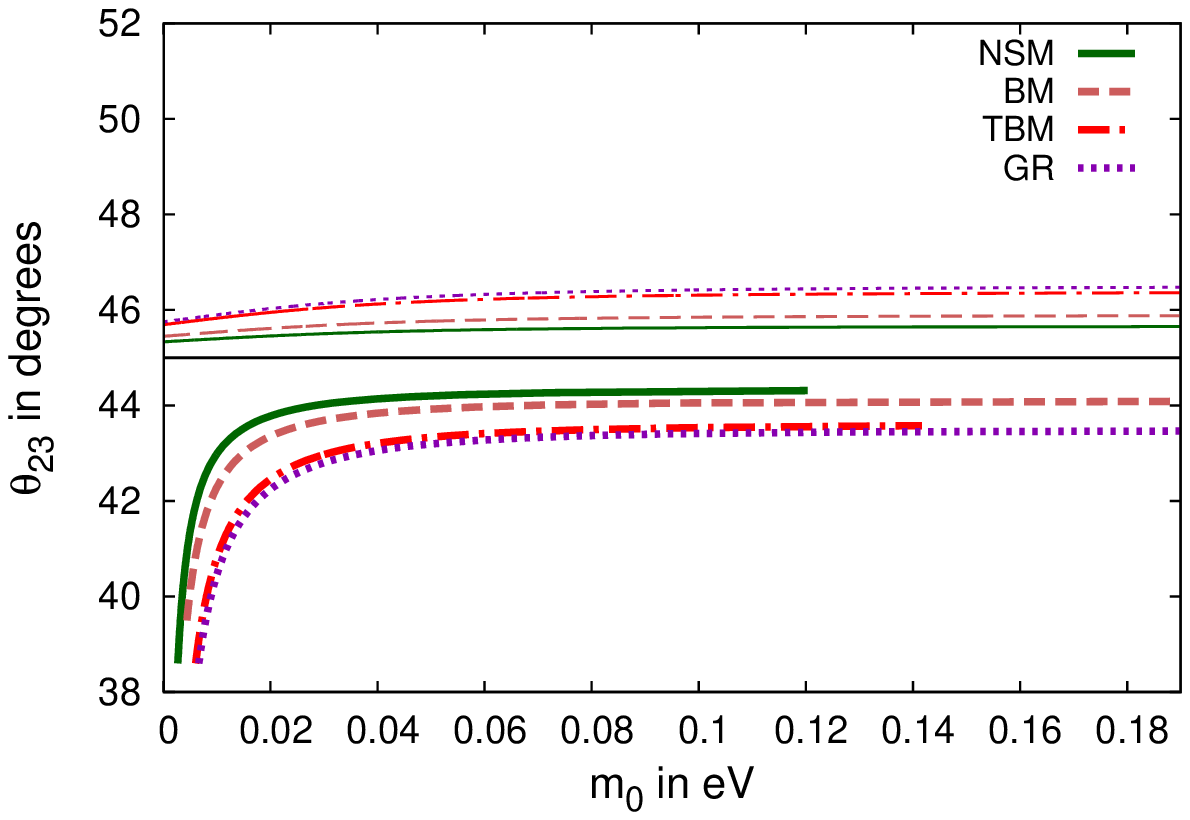}
\hspace*{0.25pt}
\includegraphics[scale=0.65,angle=0]{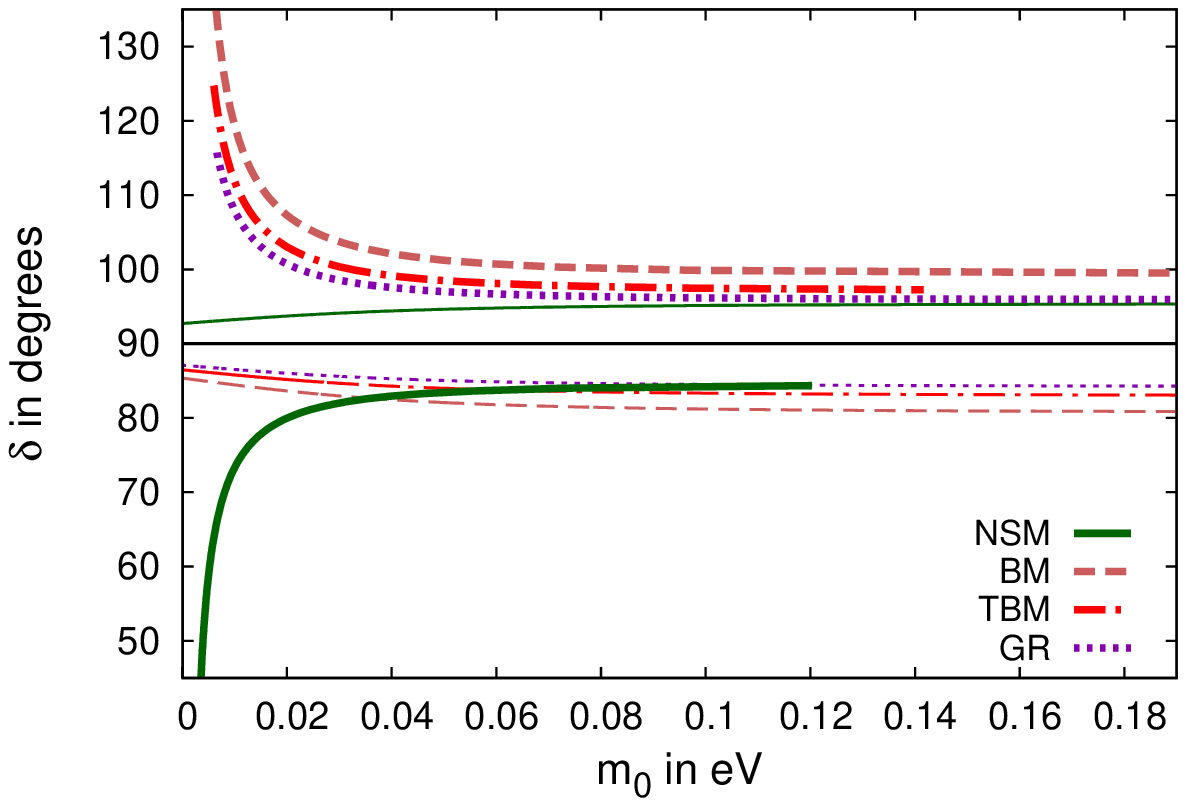}}
\caption{\em $\theta_{23}$ (left) and the CP-phase $\delta$
(right) as a function of $m_0$ from this model for different
mixing patterns when the best-fit values of the input data are
used. The NSM, BM, TBM and GR cases correspond to the green solid, pink
dashed, red dot-dashed, and violet dotted curves respectively.
Thick (thin) curves of each type indicate normal (inverted)
mass orderings. }
\label{f:CP} 
\end{center} 
\end{figure} 

As indicated in Table \ref{tlim}, $(\epsilon - \theta_{12}^0)$ always
remains in the first (fourth) quadrant for the NSM (BM, TBM, and
GR) mixing pattern. For normal ordering Eq. (\ref{cdelcomp}) then
implies that for the NSM (BM, TBM, and GR) case(s) $\delta$ lies in
the first or fourth (second or third) quadrants. For inverted
ordering of masses, $\kappa_c$ changes sign and so the quadrants
are accordingly modified. The different possibilities are
indicated in Table \ref{tabCP}. For any mixing pattern and mass
ordering there are two allowed quadrants of $\delta$ which have
$\sin \delta$ of opposite sign. Which of these is chosen
is determined by the phases $\phi_{1,2}$ through the sign of the
right-hand-side of Eq. (\ref{sdelcomp}).  As noted above,
$\phi_2$ can be in either the first or fourth quadrants and the
quadrant of $\phi_1$ is determined by the mixing pattern in such
a way that $\sin \phi_1$ can be of either sign. Thus the
the phases $\phi_1$ and $\phi_2$ can always be chosen such
that $\sin \delta$ can be of any particular sign. Therefore the two
alternate quadrants of $\delta$ for every case in Table
\ref{tabCP} are equally viable in this model.

The perturbative Type-I see-saw contribution to $\theta_{23}$ can
also be extracted from Eq. (\ref{psi3ca}). One finds:
\begin{equation}
\tan\omega = \frac{\sin\theta_{13} \cos\delta}{\tan(\epsilon - \theta_{12}^0)}
\;.
\label{th23}
\end{equation}

Recalling that Eq. (\ref{cdelcomp}) correlates $\delta$ and
$(\epsilon- \theta_{12}^0)$ through $\kappa_c$ one can readily
conclude that for all mixing patterns $~\theta_{23}$
always lies in the first (second) octant for normal (inverted)
ordering. This important conclusion from these models is shown in
Table \ref{tabCP}.

In the expression for the solar mass splitting in Eq. (\ref{solsplc1})
one can trade the factor ~$m_D^2/m_R$ in terms of $\kappa_c$ and
use Eq. (\ref{cdelcomp}) to get
\begin{equation}
\Delta m^2_{solar}
= \frac{2 m^- m_1^{(0)} \sin \theta_{13} \cos\delta \cos\epsilon}
{\sin(\epsilon - \theta_{12}^0)\ cos 2\zeta} 
\;.
\label{solsplc2}
\end{equation}

The strategy that we have followed to extract the predictions of
this model relies on utilising Eqs.  (\ref{th23}) and
(\ref{solsplc2}). We take the three mixing angles $\theta_{13}$,
$\theta_{12}$, and  $\theta_{23}$ as inputs.  With these at hand
Eq. (\ref{th23}) fixes a value of the CP-phase $\delta$. Using
these and the experimentally determined solar mass splitting one
can calculate from Eq. (\ref{solsplc2}) the combination
$m_1^{(0)} m^-$, or equivalently the variable $z$, which fixes
the lightest neutrino mass $m_0$. It might appear that arbitrarily
large values of $m_0$, and hence $m_1^{(0)} m^-$, may be admitted
by taking $\cos \delta$ to smaller and smaller values. However,
this is not the case. Experimental data require $\omega = (\pi/4 -
\theta_{23})$ to lie within determined limits. Since all other
factors have experimentally allowed
ranges, Eq. (\ref{th23}) also gives lower and upper bounds on
$\delta$. Consequently, for any mixing pattern $m_0$ lies within
a fixed range.

In the left (right) panel of Fig. \ref{f:CP} we show the mixing
angle $\theta_{23}$ (the CP phase $\delta$) as a function of the
lightest neutrino mass $m_0$ as obtained from this model for
different mixing patterns when the best-fit values of the various
measured angles and mass splittings are used. The NSM, BM, TBM
and GR correspond to the green solid, pink dashed, red
dot-dashed, and violet dotted curves respectively.  The thick
(thin) curves of each type indicate normal (inverted) mass
orderings.  Note that normal and inverted orderings are always
associated with the first and second octants of $\theta_{23}$
respectively.  For normal (inverted) ordering with the NSM mixing
pattern $\delta$ lies in the first (second) quadrant while for
the other cases it is in the second (first) quadrant. As expected, for
inverted ordering $|\delta|$ stays close to $\pi/2$  for the
entire range of $m_0$.
For normal ordering $\delta$ is  near $\pi/2$ for 
$m_0$  larger than around 0.05 eV.

Of course, as indicated in Table \ref{tabCP} if $\delta$ is a solution for
some $m_0$ then by suitably picking alternate values of the
phases $\phi_{1,2}$ which appear in $M_{\nu R}$ one can also get
a second solution with the phase $-\delta$. We have not shown
this mirror set of solutions in Fig.  \ref{f:CP}.  The  T2K
\cite{T2K} and NOVA \cite{Nova} experiments have presented data
which may be taken as a preliminary hint of normal ordering
associated with $\delta \sim -\pi/2$.  As seen from Fig.
\ref{f:CP}  this is consistent with our model, with $\delta \sim
-\pi/2$ favouring $m_0$ in the quasi-degenerate regime, i.e.,
$m_0 \geq \mathcal{O}$(0.05 eV), for normal ordering.  If this
result is confirmed by further analysis then the model will
require neutrino masses to be  in a range to which ongoing
experiments will be sensitive \cite{katrin, numass}.

The correlation between the octant of
$\theta_{23}$, the quadrant of the CP-phase $\delta$, and the
ordering of neutrino masses is a smoking-gun signal of this $S3
\times Z3$ based model.

\section{Conclusions}

In this paper we have put forward an $S3 \times Z3$ model
for neutrino mass and mixing. After assigning the flavour quantum
numbers to the leptons and the scalars we write down the most
general Lagrangian consistent with the symmetry.  Once the
symmetry is broken, the Yukawa couplings give rise to the charged
lepton masses as well as the Dirac and Majorana masses for the
left- and right-handed neutrinos.  Neutrino masses originate from
both Type-I and Type-II see-saw terms of which  the former can be
treated as a small correction.  The dominant Type-II see-saw
results in the atmospheric mass splitting, no solar splitting,
keeps $\theta_{23} = \pi/4$, and $\theta_{13} = 0$. By a choice
of the Yukawa couplings $\theta_{12}$ can be given any preferred
value.  Thus, at this level this model can accommodate any of
the much-studied tribimaximal, bimaximal, golden ratio, and `no
solar mixing' patterns. The smaller Type-I see-saw contribution
acting as a perturbation generates the solar mass splitting and
nudges the mixing angles to values  consistent with the global
fits. The octants of $\theta_{23}$ are correlated with the
neutrino mass ordering -- first (second) octant is allowed for
normal (inverted) ordering.  The model is testable through its
predictions for the CP-phase $\delta$ and from the relationships
between mixing angles and mass splittings that it entails.
Further, inverted mass ordering is correlated with a near-maximal
CP-phase $\delta$ and arbitrarily small neutrino masses are
permitted. For normal mass ordering $\delta$ can vary over a
wider range and maximality is realised in the quasi-degenerate
limit. The lightest neutrino mass must be at least a few meV in
this case.

{\bf Acknowledgements:} SP acknowledges support from CSIR, India.
AR is partially funded by  SERB Grant No. SR/S2/JCB-14/2009.

\renewcommand{\thesection}{\Alph{section}} 
\setcounter{section}{0} 
\renewcommand{\theequation}{\thesection.\arabic{equation}}  
\setcounter{equation}{0}  

\section{Appendix: Essentials of the $S3$ group}\label{AppS3}

$S3$ is a discrete group of order 6 which consists of all
permutations of three objects. It can be generated by two
elements $A$ and $B$ satisfying $A^2 = I = B^3$ and $(AB) ~(AB) =
I$. The group table is given below.

\begin{table}[tbh]
\begin{center}
\begin{tabular}{|c|c|c|c|c|c|c|}
\hline
& $I$ & $A$  & $B$ & $C$ & $D$  & $F$ \\ \hline 
$I$ & $I$ & $A$  & $B$ & $C$ & $D$  & $F$ \\ \hline
$A$ &  $A$  & $I$ & $C$ & $B$  & $F$ & $D$ \\ \hline
$F$ &  $F$  & $C$ & $I$ & $D$  & $A$ & $B$  \\ \hline
$C$ &  $C$  & $F$ & $D$ & $I$  & $B$ & $A$ \\ \hline
$D$ &  $D$  & $B$ & $A$ & $F$  & $I$ & $C$ \\ \hline
$B$ &  $B$  & $D$ & $F$ & $A$  & $C$ & $I$  \\ \hline
\end{tabular}
\end{center}
\caption{\em The group table for $S3$.}
\label{tabS3}
\end{table}

The group has two 1-dimensional representations denoted by $1$
and $1'$, and a $2$-dimensional representation. $1$ is inert
under the group  while $1^{\prime}$ changes sign under the action
of $A$. For the 2-dimensional representation a suitable choice of
matrices with the specified properties  can be readily obtained.
We choose
\begin{equation}
I = \pmatrix{1 & 0 \cr 0 & 1}\;\;,\;\; A = \pmatrix{0 & 1 \cr 1 &
0}\;\;,\;\; B = \pmatrix{\omega & 0 \cr 0 & \omega^2} \;\;,
\label{S3_21}
\end{equation}
where $\omega$ is a cube root of unity, i.e., $\omega = e^{2\pi i/3}$.
For this choice of $A$ and $B$ the remaining matrices of the
representation are:
\begin{equation}
C = \pmatrix{0 & \omega^2 \cr \omega & 0}\;\;,\;\; D = \pmatrix{0
& \omega \cr \omega^2 & 0}\;\;,\;\; F = \pmatrix{\omega^2 & 0 \cr 0
& \omega} \;\;.
\label{S3_22}
\end{equation}
The product rules for the different representations are: 
\begin{equation}
1 \times 1^\prime = 1^\prime, ~1^\prime \times 1^\prime = 1,
~{\rm and} ~2 \times 2 = 2 + 1 + 1^\prime \;\;.
\label{S3prodapp}
\end{equation}

One can see that each of the $2 \times 2$ matrices $M_{ij}$ in
eqs. (\ref{S3_21}) and (\ref{S3_22}) satisfies:
\begin{equation}
\sum_{j,l ~= 1,2} \alpha_{jl} ~M_{ij} ~M_{kl} = \alpha_{ik} \;\;,
\label{s3inv}
\end{equation}
where $\alpha_{ij} = 0$ for $i = j$ and $\alpha_{ij} = 1$ for $i \neq j$. 

If $\Phi \equiv \pmatrix {\phi_1 \cr \phi_2}$ and $\Psi \equiv
\pmatrix {\psi_1 \cr \psi_2}$ are two field
multiplets transforming under $S3$ as doublets then using eqs. (\ref{S3_21}) and (\ref{s3inv}):
\begin{equation} 
\phi_1 \psi_2 + \phi_2 \psi_1 \equiv 1 \;\; ,
\;\; \phi_1 \psi_2 - \phi_2 \psi_1 \equiv 1^\prime \;\; {\rm
and} \;\; \pmatrix{\phi_2 \psi_2 \cr \phi_1 \psi_1} \equiv 2 \;\;.
\label{S3_prod}
\end{equation}
Sometimes we have to deal with hermitian conjugate fields. Noting
the nature of the complex representation (see, for example, $B$
in eq. (\ref{S3_21})) the conjugate $S3$ doublet is $\Phi^\dagger
\equiv \pmatrix {\phi_2^\dagger \cr \phi_1^\dagger}$. As a
result,  one has in place of (\ref{S3_prod})
\begin{equation} 
\phi_2^\dagger \psi_2 + \phi_1^\dagger \psi_1 \equiv 1 \;\; ,
\;\; \phi_2^\dagger \psi_2 - \phi_1^\dagger \psi_1 \equiv 1^\prime \;\; {\rm
and} \;\; \pmatrix{\phi_1^\dagger \psi_2 \cr \phi_2^\dagger
\psi_1} \equiv 2 \;\;.
\label{S3_prod2}
\end{equation}
Eqs. (\ref{S3_prod}) and (\ref{S3_prod2}) are essential in writing
down the fermion mass matrices in Sec.
\ref{Model}.

\section{Appendix: The scalar potential and its minimum}\label{AppPot}

\setcounter{equation}{0}

As seen in Table \ref{tab1s}  this model has a rich
scalar field content. In this Appendix we write down the scalar
potential of the model keeping all these fields and derive
conditions which must be met by the coefficients of the various
terms so that
the desired {\em vev}s can be achieved. These
conditions ensure that the potential is locally minimized by this
choice.

Table \ref{tab1s} displays the behaviour of the scalar fields
under $S3 \times Z3$ besides the gauged electroweak $SU(2)_L
\times U(1)_Y$. The fields also carry a lepton number. The scalar
potential is the most general polynomial in these fields with up
to quartic terms. Our first step will be to write down the
explicit form of this potential.  Here we do not exclude any term
permitted by the symmetries.  $SU(2)_L \times U(1)_Y$ invariance
of the terms as well as the abelian lepton number and $Z3$
conservation are readily verified. It is only the $S3$ behaviour
which merits special attention.

There are a variety of scalar fields in this model, e.g.,
$SU(2)_L$ singlets, doublets, and triplets. Therefore, the scalar
potential has a large number of  terms.  For simplicity  we
choose all couplings in the potential to be real. In this
Appendix we list the potential in separate parts:  (a) those
belonging to any one $SU(2)_L$ sector, and (b) inter-sector
couplings of scalars. The $SU(2)_L$ singlet {\em vev}s, which are
responsible for the right-handed neutrino mass, are significantly
larger than those of other scalars. So, in the second category we
retain only those terms which couple the singlet fields to either
the doublet or the triplet sectors.

\subsection{$SU(2)_L$ Singlet Sector:}

The $SU(2)_L$ singlet sector comprises of two fields $\chi$ and
$\gamma$ transforming as $1(\omega)$ and $1'(\omega)$ of
$S3 ~(Z_3)$ respectively.  They have $L=0$.  The scalar potential
arising out of these is:
\begin{eqnarray}
V_{singlet}&=& m_{\chi}^2 \chi^\dagger\chi + m_{\gamma}^2 \gamma^\dagger\gamma
+\Lambda_1^s\left\{\gamma^2\chi + h.c.\right  \}
+ {\lambda_1^s \over 2} \left[\chi^\dagger\chi\right]^2
+{\lambda_2^s \over 2}\left[\gamma^\dagger\gamma\right]^2
\nonumber \\ 
&+&{\lambda_3^s \over 2}(\chi^\dagger\chi)(\gamma^\dagger\gamma)
+\lambda_4^s\left\{(\gamma^\dagger \chi)(\gamma^\dagger\chi)+ h.c.\right  \}
\;\;,
\label{Vs_s3}
\end{eqnarray} where the coefficient of the cubic term,
$\Lambda_1^s$, carries the same dimension as mass while the
$\lambda_i^s$ are dimensionless.

\subsection{$SU(2)_L$ Doublet Sector:}
The $SU(2)_L$ doublet sector of the model has two fields
$\Phi_{a,b}$ that are doublets of $S3$,  in addition to
 $\alpha$, $\beta$, and $\eta$ which are $S3$ singlets.  Among them, all
fields except $\Phi_b$ and $\alpha$ ($\in$ $\omega \ \ {\rm of}
Z_3$) are invariant under $Z_3$.
\begin{eqnarray}
V_{doublet}&=& m_{\Phi_a}^2 \Phi_a^\dagger\Phi_a
+m_{\Phi_b}^2 \Phi_b^\dagger\Phi_b
+m_{\eta}^2 \eta^\dagger\eta
+m_{\alpha}^2 \alpha^\dagger\alpha
+m_{\beta}^2 \beta^\dagger\beta
\nonumber \\ 
&+&{\lambda_1^d\over 2}\left(\Phi_a^\dagger\Phi_a\right)^2
+{\lambda_2^d\over 2}\left(\Phi_b^\dagger\Phi_b\right)^2
+{\lambda_3^d\over 2} (\Phi_a^\dagger\Phi_a)(\Phi_b^\dagger\Phi_b)
+{\lambda_4^d\over 2}(\Phi_a^\dagger\Phi_b)(\Phi_b^\dagger\Phi_a)
\nonumber \\ 
&+&{\lambda_5^d\over 2}(\Phi_a^\dagger\Phi_a)(\eta^\dagger\eta)
+{\lambda_6^d\over 2}(\Phi_a^\dagger\Phi_a)(\alpha^\dagger\alpha)
+{\lambda_7^d\over 2}(\Phi_a^\dagger\Phi_a)(\beta^\dagger\beta)
+{\lambda_8^d\over 2}(\Phi_b^\dagger\Phi_b)(\alpha^\dagger\alpha)
\nonumber \\ 
&+&{\lambda_9^d\over 2}(\Phi_b^\dagger\Phi_b)(\beta^\dagger\beta)
+{\lambda_{10}^d\over 2}(\Phi_b^\dagger\Phi_b)(\eta^\dagger\eta)
+{\lambda_{11}^d\over 2}\left(\alpha^\dagger\alpha\right)^2
+{\lambda_{12}^d\over 2}(\alpha^\dagger\alpha)(\eta^\dagger\eta)
\nonumber \\ 
&+&{\lambda_{13}^d\over 2}(\alpha^\dagger\eta)(\eta^\dagger\alpha)
+{\lambda_{14}^d\over 2}(\alpha^\dagger\alpha)(\beta^\dagger\beta)
+{\lambda_{15}^d\over 2}\left(\eta^\dagger\eta\right)^2
+{\lambda_{16}^d\over 2}(\eta^\dagger\eta)(\beta^\dagger\beta)
\nonumber \\ 
&+&\lambda_{17}^d\left\{(\Phi_a^\dagger\Phi_b)(\alpha^\dagger\eta)+ h.c.\right  \}
+{\lambda_{18}^d\over 2}\left(\beta^\dagger\beta\right)^2
\;\;.
\label{Vd_s3}
\end{eqnarray}
Leaving aside $S3$ properties for the moment, to which we
return below, out of any
$SU(2)$ doublet $\Phi$ one can construct two quartic invariants
$(\Phi^\dagger \Phi) (\Phi^\dagger \Phi)$ and $(\Phi^\dagger
\vec{\tau} \Phi) (\Phi^\dagger \vec{\tau} \Phi)$. Needless to
say, this can be generalised to the case where several distinct
$SU(2)$ doublets are involved. In order to
avoid cluttering, in Eq. (\ref{Vd_s3}) we have displayed only the
first combination for all quartic terms.  

The quartic terms involving $\lambda_1$ to $\lambda_4$ in Eq.
(\ref{Vd_s3}) are combinations of two pairs of $S3$ doublets.
Each pair can combine in
accordance to $2\times2=1+1'+2$ resulting in three terms.  The
$S3$ invariant in the potential arises from a combination of the
$1,\ \ 1', \ \ {\rm or} \ \ 2$ from one pair with the
corresponding term from the other pair.  Thus, for each such term
of four $S3$ doublets, three possible singlet combinations exist
(recall, Eq. (\ref{S3prodapp})) and
we have to keep an account of all of them.  We elaborate on this
using as an example the $\lambda_1^d$ term which actually stands
for a set of terms:
\begin{equation}
{\lambda_1^d\over 2}\left(\Phi_a^\dagger\Phi_a\right)^2
\rightarrow \lambda^d_{1_{1}}\left[(\Phi_1^\dagger\Phi_1)+(\Phi_2^\dagger\Phi_2)\right]^2
+\lambda^d_{1_{1'}}\left[(\Phi_1^\dagger\Phi_1)-(\Phi_2^\dagger\Phi_2)\right]^2
+\lambda^d_{1_{2}}\left[(\Phi_1^\dagger\Phi_2)(\Phi_2^\dagger\Phi_1)
+ (\Phi_2^\dagger\Phi_1)(\Phi_1^\dagger\Phi_2)\right].
\end{equation}
Substituting $vevs$, $\langle \Phi_1\rangle =v_1$ and 
$\langle \Phi_2\rangle =v_2$ and 
defining $\lambda^d_{1_{1}}+\lambda^d_{1_{1'}}={\lambda^d_{a_1}\over2}$
and
$2(\lambda^d_{1_{1}}-\lambda^d_{1_{1'}}+\lambda^d_{1_{2}}) =
{\lambda^d_{a_2}\over2}$
we get:
\begin{equation}
{\lambda_1^d\over 2}\left(\Phi_a^\dagger\Phi_a\right)^2
\longrightarrow {\lambda^d_{a_1}\over2}\left[(v_1^*v_1)^2+(v_2^*v_2)^2\right]
+{\lambda^d_{a_2}\over2}(v_1^*v_1)(v_2^*v_2).
\label{t6}
\end{equation}
Similarly,
\begin{equation}
{\lambda_2^d\over 2}\left(\Phi_b^\dagger\Phi_b\right)^2
\longrightarrow {\lambda^d_{b_1}\over2}\left[(v_3^*v_3)^2+(v_4^*v_4)^2\right]
+{\lambda^d_{b_2}\over2}(v_3^*v_3)(v_4^*v_4)
\label{t7}
\end{equation}
where, $\langle \Phi_3\rangle =v_3$ and 
$\langle \Phi_4\rangle =v_4$.
Further,
\begin{eqnarray}
{\lambda_3^d\over 2}\left[(\Phi_a^\dagger\Phi_a)(\Phi_b^\dagger\Phi_b)\right]
&\rightarrow& \lambda^d_{3_{1}}\left[(\Phi_1^\dagger\Phi_1+\Phi_2^\dagger\Phi_2)(\Phi_3^\dagger\Phi_3+\Phi_4^\dagger\Phi_4)\right]
+\lambda^d_{3_{1'}}\left[(\Phi_1^\dagger\Phi_1-\Phi_2^\dagger\Phi_2)(\Phi_3^\dagger\Phi_3-\Phi_4^\dagger\Phi_4)\right]
\nonumber\\
&+&\lambda^d_{3_{2}}\left[(\Phi_1^\dagger\Phi_2)(\Phi_4^\dagger\Phi_3)
+ (\Phi_2^\dagger\Phi_1)(\Phi_3^\dagger\Phi_4)\right].
\end{eqnarray}
Substituting the respective $vevs$ and defining
$\lambda^d_{3_{1}}+\lambda^d_{3_{1'}}={\lambda^d_{{ab}_1}\over 2}$,
$\lambda^d_{3_{1}}-\lambda^d_{3_{1'}}={\lambda^d_{{ab}_2}\over 2}$
and
$\lambda^d_{3_{2}}=\lambda^d_{{ab}_3}$ we get;
\begin{eqnarray}
{\lambda_3^d\over 2}\left[(\Phi_a^\dagger\Phi_a)(\Phi_b^\dagger\Phi_b)\right]
&\longrightarrow& {\lambda^d_{{ab}_1}\over 2}
\left[(v_1^*v_1)(v_3^*v_3)+(v_2^*v_2)(v_4^*v_4)\right]
+ {\lambda^d_{{ab}_2}\over 2}
\left[(v_1^*v_1)(v_4^*v_4)+(v_2^*v_2)(v_3^*v_3)\right]
\nonumber\\
&+&\lambda^d_{{ab}_3} \left[(v_1^*v_2)(v_4^*v_3)+(v_2^*v_1)(v_3^*v_4)\right].
\label{t8}
\end{eqnarray}
In a similar fashion the $\lambda_4^d$ term when expanded will lead to
\begin{eqnarray}
{\lambda_4^d\over 2}\left[(\Phi_a^\dagger\Phi_b)(\Phi_b^\dagger\Phi_a)\right]
&\longrightarrow&{ \tilde{\lambda}^d_{{ab}_1}\over 2}
\left[(v_1^*v_3)(v_3^*v_1)+(v_2^*v_4)(v_4^*v_2)\right]
+ { \tilde{\lambda}^d_{{ab}_2}\over 2}
\left[(v_1^*v_3)(v_4^*v_2)+(v_2^*v_4)(v_3^*v_1)\right]
\nonumber\\
&+& \tilde{\lambda}^d_{{ab}_3} \left[(v_1^*v_4)(v_4^*v_1)+(v_2^*v_3)(v_3^*v_2)\right].
\label{t9}
\end{eqnarray}
Adding Eqs.(\ref{t8}) and Eq.(\ref{t9}) we get:
\begin{eqnarray}
{\lambda_3^d\over 2}\left[(\Phi_a^\dagger\Phi_a)(\Phi_b^\dagger\Phi_b)\right]
+
{\lambda_4^d\over 2}\left[(\Phi_a^\dagger\Phi_b)(\Phi_b^\dagger\Phi_a)\right]
&=& { \hat{\lambda}^d_{{ab}_1}\over 2} \left[(v_1^*v_1)(v_3^*v_3)+(v_2^*v_2)(v_4^*v_4)\right]
\nonumber\\
&+&{ \hat{\lambda}^d_{{ab}_2}\over 2} \left[(v_1^*v_1)(v_4^*v_4)+(v_2^*v_2)(v_3^*v_3)\right]
\nonumber\\
&+& \hat{\lambda}^d_{{ab}_3} \left[(v_1^*v_2)(v_4^*v_3)+(v_2^*v_1)(v_3^*v_4)\right];
\label{sum_t8nt9}
\end{eqnarray}
where, ${ \hat{\lambda}^d_{{ab}_1}\over 2}\equiv
{\tilde{\lambda}^d_{{ab}_1}\over 2} + {\lambda^d_{{ab}_1}\over 2} $,
 ${ \hat{\lambda}^d_{{ab}_2}\over 2}\equiv
\tilde{\lambda}^d_{{ab}_3}  +{\lambda^d_{{ab}_2}\over 2} $
and
$\hat{\lambda}^d_{{ab}_3}\equiv
{\tilde{\lambda}^d_{{ab}_2}\over 2} +\lambda^d_{{ab}_3}  $.
Also, summing up the $\lambda_{12}^d$ and $\lambda_{13}^d$ terms
lead to ${\hat{\lambda}_{123}^d\over2} (v_\alpha^*v_\alpha)(v_\eta^*v_\eta)$,
where $\hat{\lambda}_{123}^d\equiv\lambda_{12}^d+\lambda_{13}^d$.

\subsection{$SU(2)_L$ Triplet Sector:}
Both the $SU(2)_L$ triplets present in 
our model ($\Delta_L$, $\rho_L$) that are
responsible for Majorana mass generation of the
left handed neutrinos happen to be $S3$
invariants and differ only in their $Z_3$
properties i.e., $\Delta_L(1)$ and $\rho_L(\omega)$. 
\begin{eqnarray}
V_{triplet}&=& m_{\Delta_L}^2 \Delta_L^\dagger\Delta_L
+ m_{\rho_L}^2 \rho_L^\dagger\rho_L
+ {\lambda_1^t \over 2} \left[\Delta_L^\dagger\Delta_L\right]^2
+{\lambda_2^t \over 2}\left[\rho_L^\dagger\rho_L\right]^2
+{\lambda_3^t \over 2}(\Delta_L^\dagger\Delta_L)(\rho_L^\dagger\rho_L)
\nonumber \\ 
&+&{\lambda_4^t\over 2}(\Delta_L^\dagger\rho_L)(\rho_L^\dagger\Delta_L)
+{\lambda_5^t \over 2}(\Delta_L\rho_L)(\Delta_L\rho_L)^\dagger
\;\;.
\label{Vt_s3}
\end{eqnarray}
It is noteworthy that when we write the minimized potential
in terms of the vacuum expectation values, the $\lambda_3^t$, 
$\lambda_4^t$ and $\lambda_5^t$ terms will be providing the 
same contribution as far as potential minimization is concerned.
Thus we can club these couplings together as
$\lambda_{345}^t\equiv\lambda_3^t+\lambda_4^t+
\lambda_5^t$.

\subsection{Inter-sector terms:}

So far we have listed those terms in the potential which
arise from scalars of any specific $SU(2)_L$ behaviour -- singlets,
doublets, or triplets. In addition, there can be terms which 
couple one of these sectors to another. Since the vacuum
expectation values of the singlet scalars are the largest we only
consider here the couplings of this sector to the others. The
$SU(2)_L$ triplet sector vev is very small and we drop the
doublet-triplet cross-sector couplings.

\subsubsection{$SU(2)_L$ Singlet-Doublet cross-sector:}

Couplings between the $SU(2)_L$ singlet and doublet scalars
in the potential give rise to the terms:
\begin{eqnarray}
V_{ds}&=& 
\Lambda_1^{ds}\left[(\Phi_b^\dagger\Phi_a)_{1'}\gamma + h.c.\right]
+\Lambda_2^{ds}\left[(\Phi_b^\dagger\Phi_a)_{1}\chi + h.c.\right]
+\Lambda_3^{ds}\left[(\alpha^\dagger\eta)\chi + h.c.\right]
\nonumber \\ 
&+&{\lambda_1^{ds}\over 2}(\Phi_a^\dagger\Phi_a)(\chi^\dagger\chi)
+{\lambda_2^{ds}\over 2}(\Phi_a^\dagger\Phi_a)(\gamma^\dagger\gamma)
+{\lambda_3^{ds}\over 2}(\Phi_b^\dagger\Phi_b)(\chi^\dagger\chi)
+{\lambda_4^{ds}\over 2}(\Phi_b^\dagger\Phi_b)(\gamma^\dagger\gamma)
\nonumber \\ 
&+&{\lambda_5^{ds}\over 2}(\alpha^\dagger\alpha)(\chi^\dagger\chi)
+{\lambda_6^{ds}\over 2}(\alpha^\dagger\alpha)(\gamma^\dagger\gamma)
+ {\lambda_7^{ds}\over 2}(\eta^\dagger\eta)(\chi^\dagger\chi)
+{\lambda_8^{ds}\over 2}(\eta^\dagger\eta)(\gamma^\dagger\gamma)
\nonumber \\ 
&+&\lambda_9^{ds}\left[(\Phi_a^\dagger\Phi_b)\chi^2 + h.c.\right]
+\lambda_{10}^{ds}\left[(\Phi_a^\dagger\Phi_b)\gamma^2 + h.c.\right]
+\lambda_{11}^{ds}\left[(\eta^\dagger\alpha)\chi^2 + h.c.\right]
+\lambda_{12}^{ds}\left[(\eta^\dagger\alpha)\gamma^2 + h.c.\right]
\nonumber \\ 
&+&
\lambda_{13}^{ds}\left[(\Phi_a^\dagger\Phi_b)_{1'}(\chi\gamma)+ h.c.\right]
+{\lambda_{14}^{ds}\over 2}(\beta^\dagger\beta)(\chi^\dagger\chi)
+{\lambda_{15}^{ds}\over 2}(\beta^\dagger\beta)(\gamma^\dagger\gamma)
\;\;.
\label{Vds_s3}
\end{eqnarray}

\subsubsection{$SU(2)_L$ Singlet-Triplet cross-sector:}

The terms in the potential which arise from couplings
between the $SU(2)_L$ singlet and triplet scalars are:
\begin{eqnarray}
V_{ts}&=& 
\Lambda_1^{ts}\left[(\rho_L^\dagger\Delta_L)\chi + h.c.\right]
+ {\lambda_1^{ts} \over 2} (\Delta_L^\dagger\Delta_L)(\chi^\dagger\chi)
+{\lambda_2^{ts}\over 2}(\Delta_L^\dagger\Delta_L)(\gamma^\dagger\gamma)
+{\lambda_3^{ts} \over 2}(\rho_L^\dagger\rho_L)(\chi^\dagger\chi)
\nonumber \\ 
&+&{\lambda_4^{ts}\over 2}(\rho_L^\dagger\rho_L)(\gamma^\dagger\gamma)
+\lambda_5^{ts} \left \{ (\Delta_L^\dagger\rho_L)\chi^2+ h.c.\right  \}
+\lambda_6^{ts} \left \{ (\Delta_L^\dagger\rho_L)\gamma^2+ h.c.\right  \}
\;\;.
\label{Vts_s3}
\end{eqnarray}

\subsection{The minimization conditions:}
The vevs of the scalar fields are given in Table \ref{tab1s}.
Using these:

$SU(2)_L$ singlets: $\langle \gamma^0 \rangle =u_\gamma$ and
$\langle \chi^0 \rangle =u_\chi $.

$SU(2)_L$ doublets: $\langle\Phi_a\rangle =\pmatrix{0 &v_1\cr 0&v_2}$,
$\langle\Phi_b\rangle =\pmatrix{0 &v_3\cr 0 & v_4}$
, $\langle \eta \rangle =v_\eta \pmatrix{0 & 1}$,
$\langle \alpha \rangle =v_\alpha \pmatrix{0 & 1}$ 
and $\langle \beta \rangle =v_\beta \pmatrix{1 & 0}$ .
\vskip 1pt
Recall that from the structure of the charged lepton mass matrix
Eq. (\ref{vevrat}) requires $v_2/v_1 = v_4/v_3 = A$ where the
real quantity $A = m_\mu/m_\tau$. We often also need $B \equiv (1
+ A^2)$. 

$SU(2)_L$ triplets:  $\langle \rho_L^0 \rangle
=v_\rho$ and $\langle \Delta_L^0 \rangle =v_\Delta $.


\subsubsection{$SU(2)_L$ Singlet sector:}
\begin{equation}
\frac{\partial V_{singlet}|_{min}}{\partial u_\chi^*}=0
\Rightarrow u_\chi\left[m_{\chi}^2+\lambda^s_1
u_\chi^*u_\chi\right]
+\Lambda_1^s\left(u_\gamma^*\right)^2
+ u_\gamma  \left[{\lambda_3^s \over 2}u_\chi u_\gamma^*
+2\lambda_4^su_\chi^* u_\gamma \right]=0 \;\;,
\label{MINS1}
\end{equation}
and
\begin{equation}
\frac{\partial V_{singlet}|_{min}}{\partial u_\gamma^*}=0
\Rightarrow u_\gamma\left[m_{\gamma}^2+\lambda^s_2
u_\gamma^*u_\gamma\right]
+2\Lambda_1^s\left(u_\gamma^*u_\chi^*\right)
+ u_\chi \left[{\lambda_3^s \over 2}u_\gamma u_\chi^*
+2\lambda_4^su_\gamma^* u_\chi \right]=0 \;\;.
\label{MINS2}
\end{equation}
\subsubsection{$SU(2)_L$ Doublet sector:}

Define $V_{\mathscr{D}}=V_{doublet}+V_{ds}$.
\begin{eqnarray}
\frac{\partial V_{\mathscr{D}}|_{min}}{\partial v_\alpha^*}
&=&v_\alpha\left[  m_{\alpha}^2 +{\lambda_6^d\over2}
(v_1^*v_1)B+{\lambda_8^d\over2} (v_3^*v_3)B
+\lambda_{11}^d(v_\alpha^*v_\alpha)
+{\hat{\lambda}_{123}^d\over2}(v_\eta^*v_\eta)+\lambda_{14}^d(v_\beta^*v_\beta)
\right]
\nonumber\\
&+&v_\alpha\left[ {\lambda_5^{ds}\over2}(u_\chi^*u_\chi)
+{\lambda_6^{ds}\over2}(u_\gamma^*u_\gamma)
\right]
+v_\eta\left[ \lambda_{17}^d (v_1^*v_3) B
+ \Lambda_3^{ds}u_\chi
+ \lambda_{11}^{ds}(u_\chi^*)^2
+\lambda_{12}^{ds}(u_\gamma^*)^2 \right]
=0.\nonumber\\
\label{MINT_D1}
\end{eqnarray}

\begin{eqnarray}
\frac{\partial V_{\mathscr{D}}|_{min}}{\partial v_\beta^*}
&=&v_\beta\left[ m_{\beta}^2 +{\lambda_7^d\over2}
(v_1^*v_1)B+{\lambda_9^d\over2} (v_3^*v_3)B
+{\lambda_{14}^d\over2}(v_\alpha^*v_\alpha)
+{\lambda_{16}^d\over2}(v_\eta^*v_\eta)+\lambda_{18}^d(v_\beta^*v_\beta)
\right]
\nonumber\\
&+&v_\beta\left[ {\lambda_{14}^{ds}\over2}(u_\chi^*u_\chi)
+{\lambda_{15}^{ds}\over2}(u_\gamma^*u_\gamma)\right]
=0.
\label{MINT_D2}
\end{eqnarray}

\begin{eqnarray}
\frac{\partial V_{\mathscr{D}}|_{min}}{\partial v_\eta^*}
&=&v_\eta\left[  m_{\eta}^2 +{\lambda_5^d\over2}
(v_1^*v_1)B+{\lambda_{10}^d\over2} (v_3^*v_3)B
+{\hat{\lambda}_{123}^d\over2}(v_\alpha^*v_\alpha)
+\lambda_{15}^d(v_\eta^*v_\eta)+{\lambda_{16}^d\over2}(v_\beta^*v_\beta)
\right]
\nonumber\\
&+&v_\eta\left[ {\lambda_{7}^{ds}\over2}(u_\chi^*u_\chi)
+{\lambda_{8}^{ds}\over2}(u_\gamma^*u_\gamma)
\right]
+v_\alpha\left[ \lambda_{17}^d(v_3^*v_1)B
+ \Lambda_3^{ds}u_\chi^*+\lambda_{11}^{ds}(u_\chi)^2
+\lambda_{12}^{ds}(u_\gamma)^2
\right]
\nonumber\\
&=&0.
\label{MINT_D3}
\end{eqnarray}

\begin{eqnarray}
\frac{\partial V_{\mathscr{D}}|_{min}}{\partial v_1^*}
&=&v_1\left[ m_{\Phi_a}^2 +
(v_1^*v_1)\left(\lambda_{a_1}^d+A^2{\lambda_{a_2}^d\over2}\right)
+(v_3^*v_3)\left({\hat{\lambda}_{{ab}_1}^d\over2}+A^2{\hat{\lambda}_{{ab}_2}^d\over2}
+A^2\hat{\lambda}_{{ab}_3}^d\right)
\right]
\nonumber\\
&+&v_1\left[ \left\{{\lambda_{5}^{d}\over2}(v_\eta^*v_\eta)
+{\lambda_{6}^{d}\over2}(v_\alpha^*v_\alpha)
+{\lambda_{7}^{d}\over2}(v_\beta^*v_\beta)
\right  \}
+\left\{{\lambda_{1}^{ds}\over2}(u_\chi^*u_\chi)
+{\lambda_{2}^{ds}\over2}(u_\gamma^*u_\gamma)
\right  \}
\right]
\nonumber\\
&+&v_3\left[ \left\{\lambda_{17}^d(v_\alpha^*v_\eta)
\right  \}
+\left\{
\Lambda_1^{ds}u_\gamma^*+
\Lambda_2^{ds}u_\chi^*+
\lambda_9^{ds}(u_\chi)^2
+\lambda_{10}^{ds}(u_\gamma)^2
+\lambda_{13}^{ds}(u_\chi u_\gamma)
\right  \}
\right]
\nonumber\\
&=&0.
\label{MINT_D4}
\end{eqnarray}

\begin{eqnarray}
\frac{\partial V_{\mathscr{D}}|_{min}}{\partial v_2^*}
&=&Av_1\left[  m_{\Phi_a}^2 +
(v_1^*v_1)\left(A^2\lambda_{a_1}^d+{\lambda_{a_2}^d\over2}\right)
+(v_3^*v_3)\left(A^2{\hat{\lambda}_{{ab}_1}^d\over2}+{\hat{\lambda}_{{ab}_2}^d\over2}
+\hat{\lambda}_{{ab}_3}^d\right)
\right]
\nonumber\\
&+&Av_1\left[ \left\{{\lambda_{5}^{d}\over2}(v_\eta^*v_\eta)
+{\lambda_{6}^{d}\over2}(v_\alpha^*v_\alpha)
+{\lambda_{7}^{d}\over2}(v_\beta^*v_\beta)
\right  \}
+\left\{{\lambda_{1}^{ds}\over2}(u_\chi^*u_\chi)
+{\lambda_{2}^{ds}\over2}(u_\gamma^*u_\gamma)
\right  \}
\right]
\nonumber\\
&+&Av_3\left[ \left\{\lambda_{17}^d(v_\alpha^*v_\eta)
\right  \}
+\left\{
-\Lambda_1^{ds}u_\gamma^*+
\Lambda_2^{ds}u_\chi^*+
\lambda_9^{ds}(u_\chi)^2
+\lambda_{10}^{ds}(u_\gamma)^2
-\lambda_{13}^{ds}(u_\chi u_\gamma)
\right  \}
\right]
\nonumber\\
&=&0.
\label{MINT_D5}
\end{eqnarray}
\begin{eqnarray}
\frac{\partial V_{\mathscr{D}}|_{min}}{\partial v_3^*}
&=&v_3\left[  m_{\Phi_b}^2 +
(v_3^*v_3)\left(\lambda_{b_1}^d+A^2{\lambda_{b_2}^d\over2}\right)
+(v_1^*v_1)\left({\hat{\lambda}_{{ab}_1}^d\over2}+A^2{\hat{\lambda}_{{ab}_2}^d\over2}
+A^2\hat{\lambda}_{{ab}_3}^d\right)
\right]
\nonumber\\
&+&v_3\left[ \left\{
{\lambda_{8}^{d}\over2}(v_\alpha^*v_\alpha)
+{\lambda_{9}^{d}\over2}(v_\beta^*v_\beta)
+{\lambda_{10}^{d}\over2}(v_\eta^*v_\eta)
\right  \}
+\left\{{\lambda_{3}^{ds}\over2}(u_\chi^*u_\chi)
+{\lambda_{4}^{ds}\over2}(u_\gamma^*u_\gamma)
\right  \}
\right]
\nonumber\\
&+&v_1\left[ \left\{\lambda_{17}^d(v_\eta^*v_\alpha)
\right  \}
+\left\{
\Lambda_1^{ds}u_\gamma+
\Lambda_2^{ds}u_\chi+
\lambda_9^{ds}(u_\chi^*)^2
+\lambda_{10}^{ds}(u_\gamma^*)^2
+\lambda_{13}^{ds}(u_\chi^* u_\gamma^*)
\right  \}
\right]
\nonumber\\
&=&0.
\label{MINT_D6}
\end{eqnarray}

\begin{eqnarray}
\frac{\partial V_{\mathscr{D}}|_{min}}{\partial v_4^*}
&=&Av_3\left[ m_{\Phi_b}^2 +
(v_3^*v_3)\left(A^2\lambda_{b_1}^d+{\lambda_{b_2}^d\over2}\right)
+(v_1^*v_1)\left(A^2{\hat{\lambda}_{{ab}_1}^d\over2}+{\hat{\lambda}_{{ab}_2}^d\over2}
+\hat{\lambda}_{{ab}_3}^d\right)
\right]
\nonumber\\
&+&Av_3\left[ \left\{
{\lambda_{8}^{d}\over2}(v_\alpha^*v_\alpha)
+{\lambda_{9}^{d}\over2}(v_\beta^*v_\beta)
+{\lambda_{10}^{d}\over2}(v_\eta^*v_\eta)
\right  \}
+\left\{{\lambda_{3}^{ds}\over2}(u_\chi^*u_\chi)
+{\lambda_{4}^{ds}\over2}(u_\gamma^*u_\gamma)
\right  \}
\right]
\nonumber\\
&+&Av_1\left[ \left\{\lambda_{17}^d(v_\eta^*v_\alpha)
\right  \}
+\left\{
-\Lambda_1^{ds}u_\gamma+
\Lambda_2^{ds}u_\chi+
\lambda_9^{ds}(u_\chi^*)^2
+\lambda_{10}^{ds}(u_\gamma^*)^2
-\lambda_{13}^{ds}(u_\chi^* u_\gamma^*)
\right  \}
\right]
\nonumber\\
&=&0.
\label{MINT_D7}
\end{eqnarray}

\subsubsection{$SU(2)_L$ Triplet sector:}
Define $V_{\mathscr{T}}=V_{triplet}+V_{ts}$.
\begin{eqnarray}
\frac{\partial V_{\mathscr{T}}|_{min}}{\partial v_\Delta^*}
&=&v_\Delta\left[ \left\{ m_{\Delta_L}^2 +\lambda_1^t
(v_\Delta^*v_\Delta)+{\lambda_{345}^t\over 2}(v_\rho^*v_\rho) \right  \}
+\left\{  {\lambda_1^{ts} \over 2}(u_\chi^*u_\chi)  
+ {\lambda_2^{ts} \over 2}(u_\gamma^*u_\gamma) \right  \}\right]
\nonumber\\
&+&v_\rho \left[\Lambda_1^{ts}u_\chi^*+\lambda_5^{ts}u_\chi^2
+\lambda_6^{ts}u_\gamma^2 \right]
=0.
\label{MINT1}
\end{eqnarray}

\begin{eqnarray}
\frac{\partial V_{\mathscr{T}}|_{min}}{\partial v_\rho^*}
&=&v_\rho\left[ \left\{ m_{\rho_L}^2 +\lambda_2^t
(v_\rho^*v_\rho)+{\lambda_{345}^t\over 2}(v_\Delta^*v_\Delta) \right  \}
+\left\{  {\lambda_3^{ts} \over 2}(u_\chi^*u_\chi)  
+ {\lambda_4^{ts} \over 2}(u_\gamma^*u_\gamma) \right  \}\right]
\nonumber\\
&+&v_\Delta \left[\Lambda_1^{ts}u_\chi
+\lambda_5^{ts}(u_\chi^*)^2 
+
\lambda_6^{ts}(u_\gamma^*)^2\right]
=0.
\label{MINT2}
\end{eqnarray}



\end{document}